\documentclass[pre,onecolumn,preprintnumbers,floatfix,amssymb,amsmath]{revtex4}
\usepackage{graphicx}
\usepackage{natbib}
\usepackage{subfigure}
\usepackage{multirow}
\usepackage{amssymb}
\usepackage{subfigure}
\usepackage{amsmath}
\usepackage{multirow}

\begin{document}
\title{Equilibrium solutions of a sessile drop partially covered by another liquid}
\author{P. D. Ravazzoli, A. G. Gonz{\'a}lez, J. A. Diez}
\affiliation{Instituto de F\'{\i}sica Arroyo Seco, Universidad Nacional del Centro de la Provincia de Buenos Aires, and CIFICEN-CONICET-CICPBA, Pinto 399, 7000, Tandil, Argentina}

\begin{abstract}
We study the equilibrium solutions of a sessile drop on top of a horizontal substrate when it is partially covered by another inmiscible liquid, so that part of the drop is in contact with a third fluid (typically, air). The shapes of the interfaces are obtained by numerical integration of the differential equations resulting from the application of the Neumann's and Young's laws at the three fluids junction (triple point) and the two fluids--solid contact line (at the substrate), as well as no--flow boundary conditions at the wall of the container under axial symmetry. The formulation shows that the solutions are parametrized by four dimensionless variables, which stand for: i) the volume of the external liquid in the container,  ii) two  ratios of interfacial tensions, and iii) the contact angle of the liquid drop at the substrate. In this parameters space, we find solutions that consist of three interfaces: one which connects the substrate with the triple point, and two that connect this point with the symmetry axis and the wall. We find that the shape of the first interface strongly depend on the height of the external liquid, while the contact angle and the two ratios of surface tensions do not introduce qualitative changes. In particular, the solution can show one or more necks, or even no neck at all, depending on the value of the effective height of the external fluid. An energetic analysis allows to predict the breakups of the necks, which could give place to separated systems formed by a sessile drop and a floating lens, plus eventual spherical droplets in between (depending on the number of breakups).

\end{abstract}

\maketitle

\section{Introduction}
The knowledge of a drop shape when it is surrounded by other inmiscible liquids and/or in contact with solid bodies is a subject that has always raised interest in fluids mechanic. A non exhaustive list of examples could include the shapes of a sessile drop under different wettability conditions on a solid substrate~\cite{deGennes1985} as well as on another liquid~\cite{Seeto1983,Wyart1993,Takamura2012,Ravazzoli_PRF_2020}, or that of a liquid bridge that connects two solid surfaces such as disks or rings~\cite{Borkar1991,vanHonschoten2010,Orr1975,Lowry1995,Slobozhanin1997}. The solution of these basic problems, even if they seem to have only an academic appeal, can lead to the development of new techniques to manipulate tiny droplets with application in several of fields of industry~\cite{Kumar2015,Lohse2015,Gates2005}.

All of the above mentioned configurations involve two fluids (usually a liquid and a gas) and a solid substrate. The inclusion of another liquid phase has also been of interest in the literature~\cite{Neeson2012,Zhang2016}, as it is the case of compound or multiphase drops. These ones are comprised of two (or more) immiscible fluid drops that share an interface with one another, surrounded by a third, mutually immiscible fluid. Such drops exist in several different areas, such as multiphase processing, biological interactions within cells and atmospheric chemistry. In some cases, the solid substrate does not play a crucial role in determining the equilibrium shapes, while in other cases one has a four--phases problem, namely, two liquids, one gas and a solid. The problem we consider here belongs to this latter case.

\begin{figure}[ht]
\includegraphics[width=0.5\linewidth]{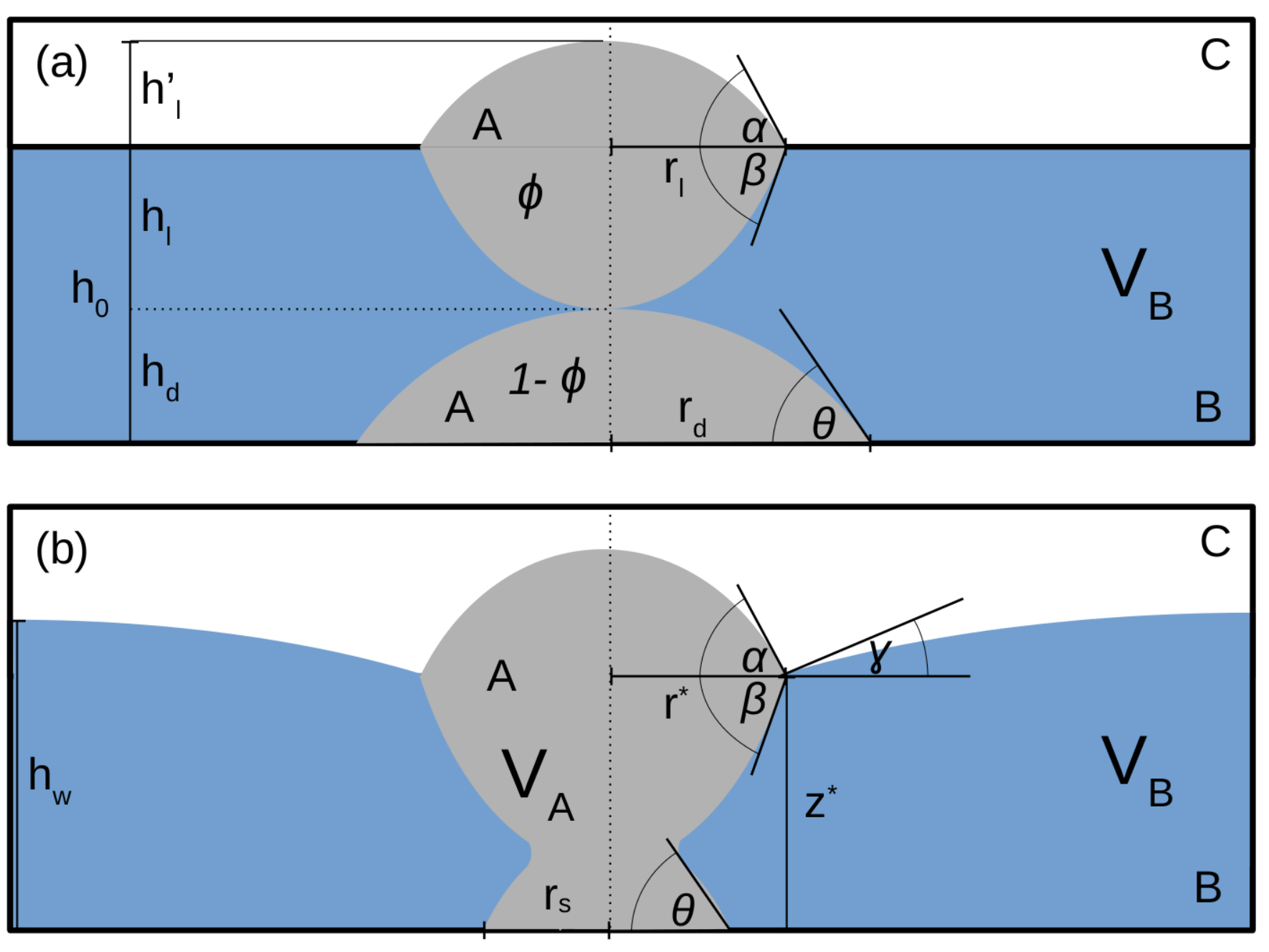}
\caption{Schemes of the initial and final equilibrium states: (a) Initial state where a sessile drop of volume $(1-\phi) \mathcal{V}_A$ rests on a substrate while a liquid lens of volume $\phi \mathcal{V}_A$ touches the drop as the liquid $B$ level reaches the value $h_0$. (b) Equilibrium solution to be found.}
\label{fig:schemeInit}
\end{figure}

In particular, we are concerned with the possible final equilibrium configuration attainable, under capillary forces but neglecting gravity, when \emph{a lens of liquid $A$ resting on a film of liquid $B$ reaches the top of a sessile drop of liquid $A$ sumerged in the film and supported by a solid substrate}  (see Fig.~\ref{fig:schemeInit}(a)). Since both the lens and the drop are formed by liquid $A$ with total volume ${\mathcal V}_A$, the former contains the fraction $\phi$ of that volume while the latter contains the fraction $(1-\phi)$ with $0<\phi<1$. The extreme cases $\phi=0$ and $\phi=1$ correspond to a unique drop (no lens) and a unique lens (no drop), respectively. The liquid $B$, of volume ${\mathcal{V}_B}$, which supports the lens and surrounds the drop, is immiscible, while fluid $C$ (on top of the lens and liquid $B$) can be another immiscible liquid or simply another gas, like air. Here, we focus on finding the static equilibrium configuration (see Fig.~\ref{fig:schemeInit}(b)), and not on the dynamics of the evolution towards it. 
Even if $\mathcal{V}_B \gg \mathcal{V}_A$, we consider that $\mathcal{V}_B$ is finite since it corresponds to the liquid volume in a cylindrical vessel of finite radius, $r_w$. We will show that the value of $\mathcal{V}_B$ is important to determine the equilibrium solution.

A main issue of concern regarding the equilibrium solutions is their stability. If their unstability leads to breakups, then one could expect that the final configuration might reach a separated system  formed by at least two bodies of liquid $A$: a new sessile drop on top of the substrate and another liquid lens at the interface $BC$. In order to gain some insight regarding this instability issue, we compare the surface energies for each type of the unbroken equilibrium solutions with that of the broken configuration. 

This problem can be seen as an extension of the ordinary sessile drop, in which the surrounding media includes now a new interface ($BC$ interface, see Fig.~\ref{fig:schemeInit}(b)). Also, since one could finally obtain a liquid lens formed by part of the sessile drop volume, the present problem can also be considered as an extension of the liquid lens problem where the substrate is not infinitely far away from the floating drop, but it interacts with the lens by direct contact (see Fig.~\ref{fig:schemeInit}(b)). In any case, we deal here with a four phases problem since we consider not only three fluids (two liquids and air), but also the presence of the solid substrate. Moreover, it could be considered a five phases problem if we add the normal contact of interface $BC$ with the vertical wall of the vessel radius at $r_w$.

The paper is organized as follows: In Section~\ref{sec:problem}, we formulate the basic problem in a nondimensional way, and we describe the numerical scheme developed to calculate the axisymmetric equilibrium solutions. We show that they present a certain number of regions with local minima radii, called necks. In Section~\ref{sec:Bridges}, we analyze the external fluid $B$ level effects on the different types of solutions, mainly related to their number of necks. Since the equilibrium solutions might break up at these necks, we focus on their energetic content to assess their stability. Then, we evaluate the surface energies associated with each type of solution, and compare them with those of the separated systems. These are formed basically by a sessile drop and a floating lens and eventually also by spherical droplets in between resulting from the breakups. In Section~\ref{sec:ContactAngle}, we analyze the effects related with the wettability of the substrate by considering a wide range of contact angles, and Section~\ref{sec:eta} deals with the effects of the ratios of the surface tensions. Finally, in Section~\ref{sec:conclu}, we summarize the results and discuss their implications. 

\section{Formulation of the problem}
\label{sec:problem}

\begin{center}
\begin{figure}[ht]
\includegraphics[width=0.6\linewidth]{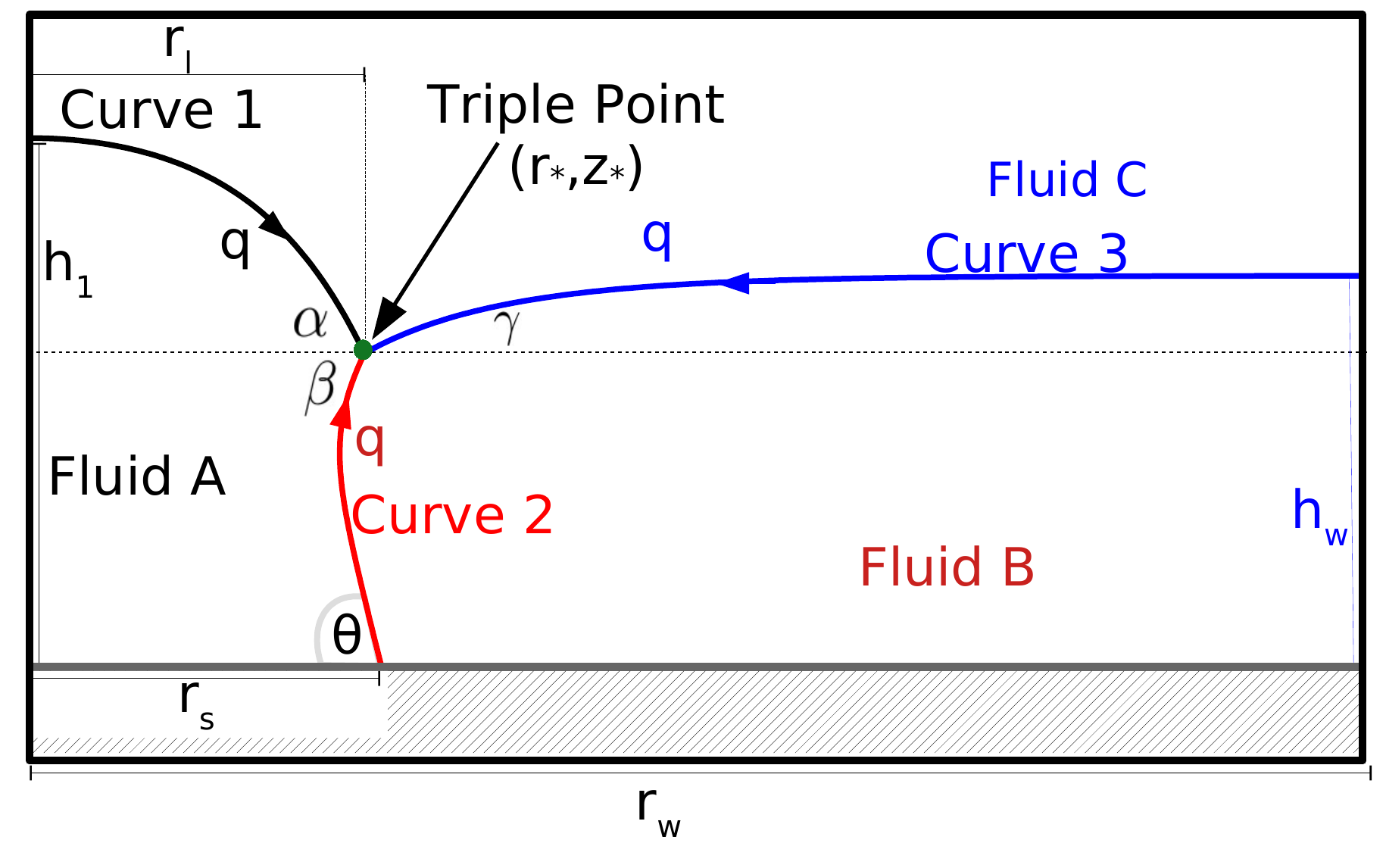}
\caption{Scheme of the equilibrium solution: A liquid bridge (curve $2$) connects the triple point $(r_\ast,z_\ast)$ with the solid substrate where the contact angle is $\theta$. Curve $1$ encloses the upper part of fluid $A$ and curve $3$ incloses the total volume of fluid $B$. The arrows indicate the increasing direction of the arc length, $q$. The angle $\alpha$, $\beta$ and $\gamma$ are positive in this scheme.}
\label{fig:bridgeScheme}
\end{figure}
\end{center}

In order to obtain the equilibrium solution, we consider axial symmetry so that each surface, $S$, can be reduced to a single curve in the $r$--$z$ plane. We assume that the equilibrium solution adopts the typical shape as depicted in Fig.~\ref{fig:bridgeScheme}. Therefore, curve $1$ corresponds to $S_{AC}$, curve $2$ to $S_{AB}$ and curve $3$ to $S_{BC}$. At the solid substrate, we consider that curve $2$ makes a contact angle $\theta$, which determines the degree of wettability of the solid by liquid $A$ when surrounded by liquid $B$.  

At equilibrium, the angles $(\alpha,\beta, \gamma)$ at the triple point ($A$--$B$--$C$) of coordinates $(r_\ast,z_\ast)$ (see Fig.~\ref{fig:bridgeScheme}) must satisfy the Neumann conditions
\begin{equation}
\alpha + \beta = \arccos \left( \frac{\sigma_{1}^2 - \sigma_{2}^2 - \sigma_{3}^2}{2 \sigma_{3} \sigma_{2}}\right), \quad
\alpha + \gamma =\pi - \arccos \left( \frac{\sigma_{2}^2 - \sigma_{3}^2 - \sigma_{1}^2}{2 \sigma_{3} \sigma_{1}}\right),
\label{eq:Neumann}
\end{equation} 
where the subindexes in the surface tensions $\sigma$'s account for the corresponding interface (the scheme in  Fig.~\ref{fig:bridgeScheme} shows positive values of these angles). In the case of a lens floating on a liquid of infinite extension (both laterally and in depth), we have $\gamma=0$, as it is the case when the lens does not touch the substrate~\cite{Ravazzoli_PRF_2020}. Once the lens is in contact with the sessile drop on the substrate, the interface $BC$ is not longer flat and then it must be $\gamma \neq 0$ at the triple point. At the same time, the height of this interface at the lateral wall, $r= r_w$, changes to $h_w \neq h_0$. Even if we will consider that $r_w$ is much larger than the drop or lens radii, $r_d$ or $r_l$, respectively, its finite value defines the amount of fluid $B$ in the vessel, ${\cal V}_B$. For simplicity, we will assume that the contact angle of the interface $BC$ with the wall is fixed and equal to $\pi/2$. In other words, we do not consider relevant the wettability of the wall, otherwise, the problem would be one of five phases instead of four.     
It can be shown that the relations in Eq.~(\ref{eq:Neumann}) can be satisfied only if the surface tensions satisfy~\cite{Ravazzoli_PRF_2020}
\begin{equation}
-2 \min \left( \sigma_1,\sigma_2 \right) < S = \sigma_3 -\sigma_1 - \sigma_2< 0,
\label{eq:S}
\end{equation}
where $S$ is the spreading parameter of $A$ laying on $B$ and surrounded by $C$. The minimum value of $S$ assures that the angles $(\alpha, \beta, \gamma)$ can actually be computed, while its maximum value ($S=0$) implies that these angles are null (i.e., $A$ spreads indefinitely over $B$). Then, under these constraints, the force balance at the triple point in the horizontal and vertical directions are:
\begin{subequations}
\begin{equation}
\sigma_{1} \cos \alpha + \sigma_{2} \cos \beta - \sigma_{3} \cos \gamma = 0, 
\label{eq:neumaannCos}
\end{equation}
\begin{equation}
\sigma_{1} \sin \alpha - \sigma_{2} \sin \beta + \sigma_{3} \sin \gamma = 0.
\label{eq:neumaannSin}
\end{equation}
\end{subequations}
Each curve, representing the interfaces, will be defined by the Laplace pressure jump condition:
\begin{equation}
\Delta p_i=\sigma_i \kappa_i = \sigma_i \left( \frac{1}{{\mathcal R}_{a,i}} + \frac{1}{{\mathcal R}_{b,i}} \right),
\label{eq:Lapl}
\end{equation}
where $\kappa_{i}$ is the curvature of the $i$--th curve and ${\mathcal R}_{a,i}$ and ${\mathcal R}_{b,i}$ are the curvature radii in the $r$--$z$ plane and in the perpendicular plane that contains the normal to the curve $i$ at the point, respectively. 
Besides, the pressure equilibrium implies that the  sum of the pressure jumps, $\Delta p_i=\sigma_i \kappa_i$ at the three interfaces, must be zero, so that
\begin{equation}
 \sigma_{1} \kappa_{1} + \sigma_{2} \kappa_{2} + \sigma_{3} \kappa_{3} = 0.
\label{eq:balPres}
\end{equation}
Note that these curvatures are constant along the curves, since gravity effects are neglected.

In order to formulate the problem in a \emph{dimensionless} form, we define its characteristic length as the radius of a spherical drop of volume $\mathcal{V}_A$ ,
\begin{equation}
{\mathcal R}_0 = \left( \frac{3 \mathcal{V}_A}{4 \pi} \right)^{1/3}.
\label{eq:R0}
\end{equation}
Therefore, the \emph{dimensionless} volume of liquid $A$ is $V_A=4\pi/3$. 

The \emph{dimensionless} equations that govern the Laplace (capillary) pressure jumps along the curves $1$, $2$ and $3$ in parametric form are~\cite{Burton2010,Ravazzoli_PRF_2020} 
\begin{subequations}
\label{eq:rizi}
\begin{equation}
r_i''(q_i) = \,z_i'(q_i) \left[ \frac{z_i'(q_i)}{r_i(q_i)} - L_i \Delta P_i \right],
\end{equation}
\begin{equation}
z_i''(q_i) = -r_i'(q_i) \left[ \frac{z_i'(q_i)}{r_i(q_i)} - L_i \Delta P_i \right],
\end{equation}
\end{subequations}
where $(r,z)$ are the \emph{dimensionless} cylindrical coordinates in units of ${\mathcal R}_0$ and $\Delta P_i=({\mathcal R}_0\sigma_i) \Delta p_i$ are the dimensionless pressure jumps. Here, $q_i=s/L_i$ is the scaled arc length along the curves, where $L_i$ is the dimensionless total arc length of the $i$--th curve. Since we define $q_i=1$ at the triple point where all curves meet and end up, it corresponds $q_i=0$ to the point where they begin: $(0,h_1)$ for curve $1$, $(r_d,0)$ for curve $2$, and $(r_w,h_w)$ for curve $3$. Consequently, we have  
\begin{equation}
L_i^2 = r'_i(q_i)^2 + z'_i(q_i)^2 = const.
\end{equation} 
In Table~\ref{tab:bc}, we summarize the \emph{twelve} boundary conditions needed for the integration of the \emph{six} second order differential equations in Eq.~(\ref{eq:rizi}). Note, however, that these equations have \emph{six}  unknowns, namely, three $\Delta P_i$'s and three $L_i$'s. Moreover, the conditions listed in Table~\ref{tab:bc} add \emph{three} additional unknowns, namely, $r_d$, $h_1$ and $h_w$ (the contact angle $\theta$ is a given parameter). Therefore, we have \emph{nine} unknows in total, namely, 
\begin{equation}
{\cal U}=(\Delta P_1, \Delta P_2, \Delta P_3, L_1, L_2, L_3, r_d, h_w, h_1),
\label{eq:U}
\end{equation}
 so that we need \emph{nine} constraints to solve the problem. 
\begin{table}[ht]
\begin{tabular}{|c|c|c|c|}
\hline \hline
& Curve~$1$ & Curve~$2$ & Curve~$3$ \\ \hline
$r(q_i=0)$ & 0 & $r_d$ & $r_w$   \\ \hline
$r'(q_i=0)$ & $L_1$ & $-L_2 \cos \theta$ & $-L_3$  \\ \hline
$z(q_i=0)$ & $h_1$ & 0 & $h_w$  \\ \hline
$z'(q_i=0)$ & 0 & $L_2 \sin \theta$ & 0 \\ \hline \hline
\end{tabular}
\caption{Boundary conditions for the \emph{six} equations in Eq.~(\ref{eq:rizi}).}
\label{tab:bc}
\end{table}

Naturally, \emph{three} constraints are given by Eqs.~(\ref{eq:neumaannCos}), (\ref{eq:neumaannSin}) and (\ref{eq:balPres}), which must be written in \emph{dimensionless} form. In fact, by considering $\sigma_{1}$ as a reference surface tension, $\sigma_{ref}=\sigma_1$, we define the dimensionless ratios as
\begin{equation}
\eta = \frac{\sigma_2}{\sigma_1}, \quad
\zeta = \frac{1}{2} \left( 1 + \eta -\frac{\sigma_3}{\sigma_1} \right)
\label{eq:eta_zeta}
\end{equation}
Thus, the restrictions in Eq.~(\ref{eq:S}) imply
\begin{equation}
\eta >0, \qquad 0<\zeta<\min(1,\eta).
\end{equation}
Therefore, the first \emph{three} constraints are given by the \emph{dimensionless} version of the force and pressure balances,
\begin{subequations}
\begin{equation}
\cos \alpha + \eta \cos \beta - \left( 1 + \eta - 2 \zeta \right) \cos \gamma = 0, 
\label{eq:neumaannCosAd}
\end{equation}
\begin{equation}
\sin \alpha - \eta \sin \beta + \left( 1 + \eta - 2 \zeta \right) \sin \gamma = 0,
\label{eq:neumaannsinAd}
\end{equation}
\begin{equation}
\kappa_{1} + \eta \kappa_{2} + \left( 1 + \eta - 2 \zeta \right) \kappa_{3} = 0,
\label{eq:balPresAd}
\end{equation}
\label{eq:NeumannPresAd}%
\end{subequations}
where the $\kappa_{i}$'s are now in units of $\mathcal{R}_0$.

Other \emph{four} constraints come from the fact that all three curves must meet at the triple point  of coordinates $(r_\ast,z_\ast)$. Thus, we have
\begin{subequations}
\begin{align}
r_\ast &= r_1 (q_1=1) = r_2 (q_2=1) = r_3 (q_3=1), \\
z_\ast &= z_1 (q_1=1) = z_2 (q_2=1) = z_3 (q_3=1).
\end{align}
\label{eq:ptCond}%
\end{subequations}

The remaining \emph{two} constraints are given by the volume conservation of both $A$ and $B$. They  are of integral type, and can be written in \emph{dimensionless} form as
\begin{subequations}
\label{eq:consVol}
\begin{align}
V_A &=  \pi \left( \int_0^1 r_2^2 z'_2 dq_2 + \int_0^1 r_1^2 z'_1 dq_1 \right)=\frac{4\pi}{3},\\
V_B &=\pi r_w^2 h_w - \pi \left(\int_0^1 r_3^2 \vert z'_3 \vert dq_3 + \int_0^1 r_2^2 \vert z'_2 \vert dq_2 \right),
\label{eq:consVolB}
\end{align}
\end{subequations}
where $V_A$ and $V_B$ are in units of ${\mathcal R}_0^3$. In summary, the \emph{nine} constraints to calculate the \emph{nine} unknowns are given by Eqs.~(\ref{eq:NeumannPresAd}), (\ref{eq:ptCond}) and (\ref{eq:consVol}), while the parameters $\eta$, $\zeta$, $\theta$, $V_B$ and $r_w$ are given.

\subsection{Numerical procedure}
\label{sec:NumProc}

\begin{figure}[ht]
\includegraphics[width=0.6\linewidth]{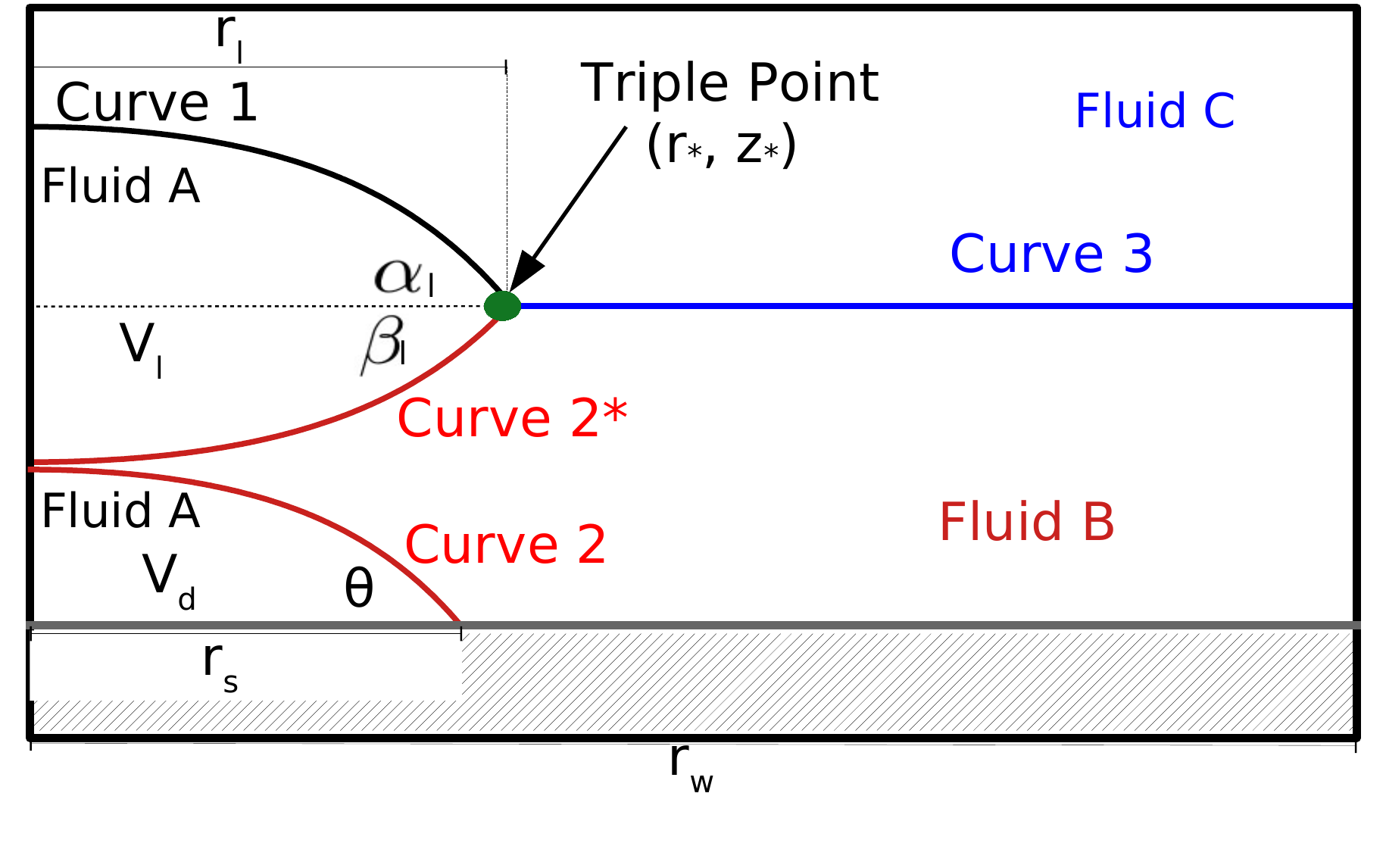}
\caption{Scheme of the \emph{guess} solution: A liquid lens and sessile drop.}
\label{fig:lensdropScheme}
\end{figure}

Here, we develop an algorithm that numerically integrates Eq.~(\ref{eq:rizi}), under the boundary conditions in Table~\ref{tab:bc}, within an iteration procedure that tries to satisfy the nine constraints to look for a globally convergent solution. To do so, it is required to start the iteration with a \emph{guess} solution. We have identified that a suitable \emph{guess} solution is composed by a \emph{floating lens} in contact with a \emph{sessile drop} on the substrate, both formed by $A$, for a given volume $V_B$ in the vessel (Fig.~\ref{fig:lensdropScheme}). Thus, we consider that the lens and drop have volume $V_l=\phi_0 V_A$ and $V_d=(1-\phi_0) V_A$, respectively, with $0 <\phi_0 <1$. 

This guess solution can be analytically described as follows. 

Firstly, the shape of the upper lens can be obtained from the simultaneous solution of Eqs.~(\ref{eq:Lapl}) and (\ref{eq:balPres}). In this case, we have $\kappa_3=0$ because the interface $BC$ extends to infinity where the curvature is zero, so that $\Delta P_3=0$. Consequently, $\gamma=0$ and from Eqs.~(\ref{eq:Neumann}) and (\ref{eq:eta_zeta}), we determine $\alpha_l$ and $\beta_l$ as 
\begin{equation}
\begin{aligned}
\alpha_l &= \pi - \arccos \left[\frac{\eta^2-1-\left(1+\eta -2 \zeta \right)^2}{2 \left(1+\eta -2 \zeta \right)}\right]\\
\beta_l &= \arccos \left[ 1 + \frac{2 \zeta \left( \zeta - 1- \eta \right)}{\eta}\right] - \alpha_l.
\end{aligned}
\end{equation}
Since $\Delta P_2=-\Delta P_1$ and $\sigma_2=\eta \sigma_1$, the lens is composed of two inverted spherical caps of different curvature that meet at the triple point, the lens radius $r_l$~\cite{Ravazzoli_PRF_2020} (see Fig.~\ref{fig:lensdropScheme}),
\begin{equation}
r_{l} = 2 \, \left( \phi_0 V_A \right)^{1/3} \left(3 \tan \frac{\alpha_l}{2} + 3 \tan \frac{\beta_l}{2} + \tan^3 \frac{\alpha_l}{2} + \tan^3 \frac{\beta_l}{2}\right)^{-1/3}.
\end{equation}

Secondly, the solution of the sessile drop on the solid substrate is given by a single spherical cap of radius
\begin{equation}
r_{d} = 2 \,\left[ (1-\phi_0 ) V_A \right]^{1/3} \left( 3 \tan \frac{\theta}{2} +  \tan^3 \frac{\theta}{2}\right)^{-1/3},
\label{eq:rd}
\end{equation}
which is a special case of $r_l$ with $\alpha_l=\theta$ and $\beta_l=0$ and the corresponding volume. 
Since $V_A=4\pi/3$, both the radii $r_d$ and $r_l$ depend on $\phi$. 

Therefore, the \emph{guess} values of our nine unknowns are taken from this configuration for a given $\phi_0$ with the only difference that $\Delta P_3$ is not zero, but a small number $\epsilon$:
\begin{equation}
\begin{aligned}
\Delta P_{1} &= - \frac{\sin \alpha_l}{r_l} \qquad & L_{1} = r_l \frac{\alpha_l}{\sin \alpha_l} \qquad & r_s = r_d\\
\Delta P_{2} &= \eta \frac{\sin \beta_l}{r_l} \qquad & L_{2} = r_l \frac{\beta_l}{\sin \beta_l} \qquad & h_w = h_0\\
\Delta P_{3} &=\epsilon \qquad \qquad & L_{3} = r_w-r_l \qquad & h_1 = h_{d}+ h_{l} + h_l^\prime.
\end{aligned}
\label{eq:guess_values}
\end{equation}

\begin{figure}[ht]
\includegraphics[width=0.5\linewidth]{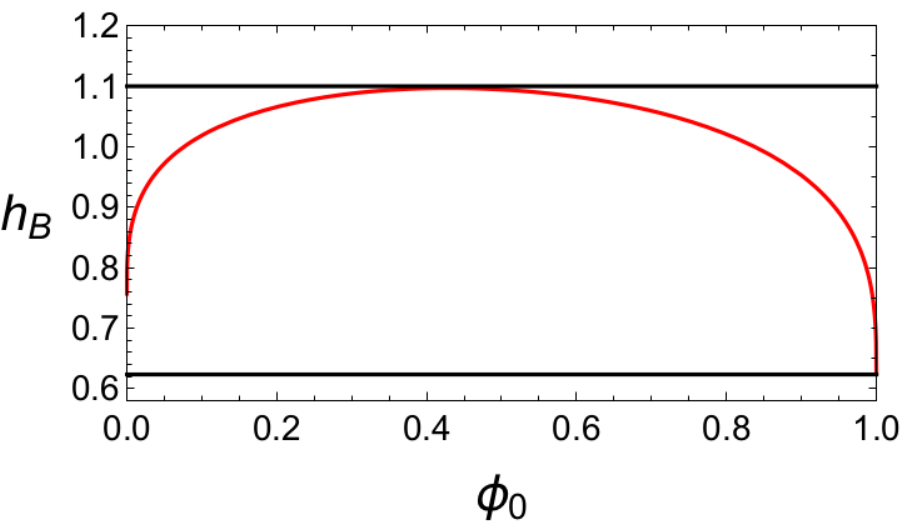}
\caption{Equivalent thicknes $V_B$, $h_B$, as function of $\phi_0$ for the configuration \emph{lens + drop} when $h_0=h_l+h_d$ (see Fig.~\ref{fig:schemeInit}(a)).}
\label{fig:volBIni}
\end{figure}

Typically, we use $\epsilon=10^{-5}$ and $r_w=30$. Then, the problem is solved for a given volume $V_B$, or its equivalent thickness
\begin{equation}
h_B = \frac{V_B}{\pi r_{w}^2},
\label{eq:hB}
\end{equation}
whose dependance on $\phi_0$ is shown in Fig.~\ref{fig:volBIni}.

Thus, we can start the integration of Eq.~(\ref{eq:rizi}) from $q_i=0$ to $q_i=1$ including the boundary conditions and the volume constraints in Eq.~(\ref{eq:consVol}). Through an iterative procedure on the nine parameters in ${\cal U}$ (see Eq.~(\ref{eq:U})), we are able to obtain a new solution along with the corresponding values of the nine converged parameters. This solution will have a shape as depicted in Fig.~\ref{fig:bridgeScheme}, 

After one solution is reached, we use the converged list ${\cal U}$ for the guess values for a new geometrical configuration with a modified value of $h_B \rightarrow h_B \pm \delta h_B$ (typically, $\delta h_B = 0.015$, which implies $\delta V_B \approx 10 V_A=40\pi/3$). Thus, the $n$--th solution corresponds to the equivalent thickness
\begin{equation}
h_B (n)=h_B(n-1)\pm \delta h_B,
\label{eq:hBn}
\end{equation} 
where $n \geq 2$ stands for the iteration step. Note that, by using this scheme, it is possible to obtain solutions for volumes $V_B$ (and corresponding $h_B$) out of the interval shown in Fig.~\ref{fig:volBIni} ($h_B=(0.622,1.096)$). When $h_B$ is within that interval, it is convenient to use directly the corresponding value of $\phi_0$. In this case, the parameters $\eta$, $\zeta$ and $\theta$ remain invariable.

Analogously, when the contact angle $\theta$ is varied for given $\eta$, $\zeta$ and $h_B$, we employ a similar methodology by defining
\begin{equation}
\theta (n)=\theta(n-1)\pm \delta \theta,
\end{equation} 
where, typically, $\delta \theta = 1^\circ$.

The equilibrium solution depends on the physical properties of the liquids, their wetabilitty with the solid bottom and the liquid volume $B$ (or, equivalently, $h_B$ as defined by Eq.~(\ref{eq:hB})), that is, on the list of parameters ${\cal F}=(\eta, \zeta, \theta, h_B)$. In the following sections,  we present a parametric study of the solution by varying $h_B$, $\theta$, or $\eta$ with the other parameters remaining fixed. Note that varying $\eta$ is equivalent to changing $\zeta$, since these parameters only modify the ratii of the surface tensions (see Eq.~(\ref{eq:eta_zeta})). For varying $\eta$, we use
\begin{equation}
\eta (n)=\eta(n-1)\pm \delta \eta,
\end{equation} 
with $\delta \eta=0.05$.

\section{Effects of $h_B$}
\label{sec:Bridges}

In this section, we focus on the effects of $h_B$, keeping fixed the liquids properties and their wettability with the bottom, by choosing $\eta=0.8$, $\zeta=0.2$ and $\theta=45^{\circ}$. Thus, we have three fluids with the following ratii of surface tensions: $\sigma_2=0.8 \,\sigma_1$ and $\sigma_3=1.4 \,\sigma_1$.

\begin{figure}[ht]
\subfigure [$h_B=0.5082$] {\includegraphics[width=0.24\linewidth]{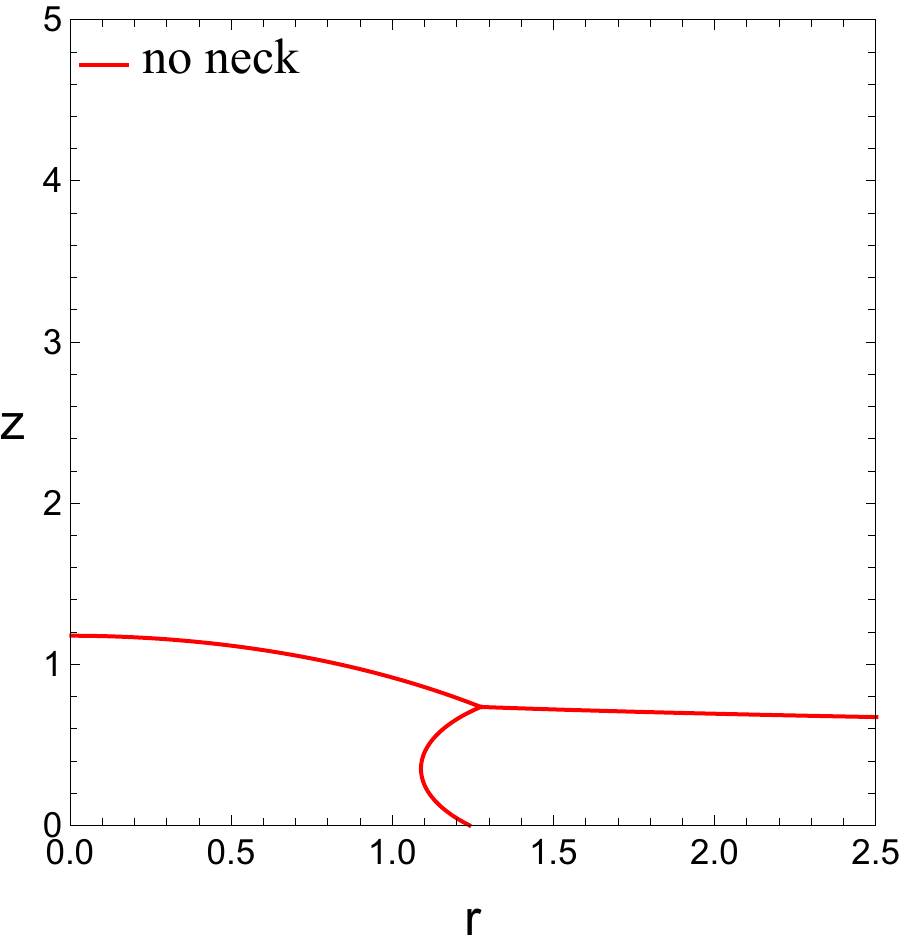}}
\subfigure [$h_B=1.0119$] {\includegraphics[width=0.24\linewidth]{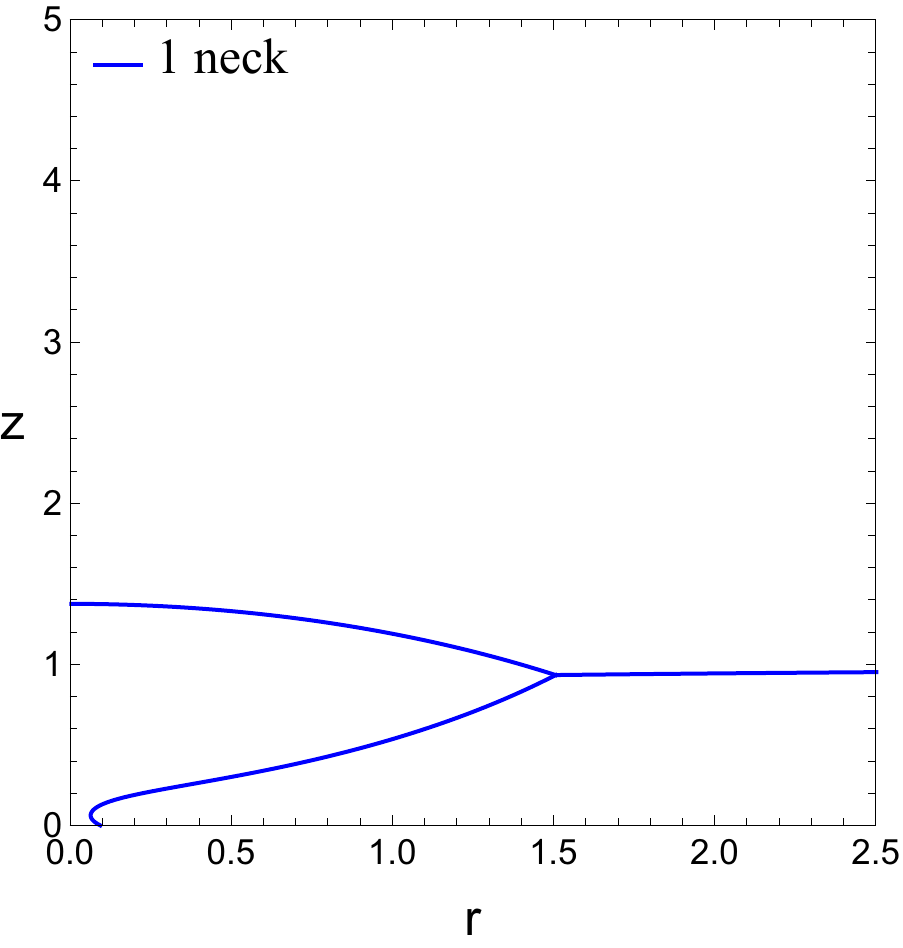}}
\subfigure [$h_B=2.9798$] {\includegraphics[width=0.24\linewidth]{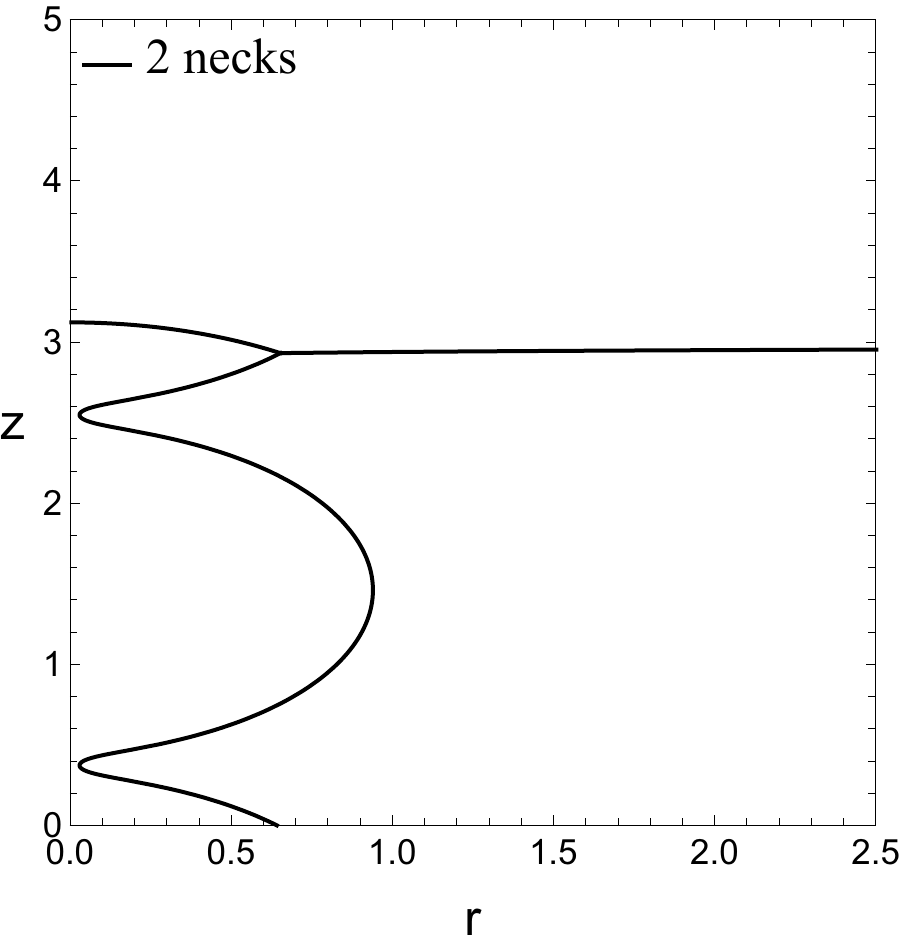}}
\subfigure [$h_B=4.0054$] {\includegraphics[width=0.24\linewidth]{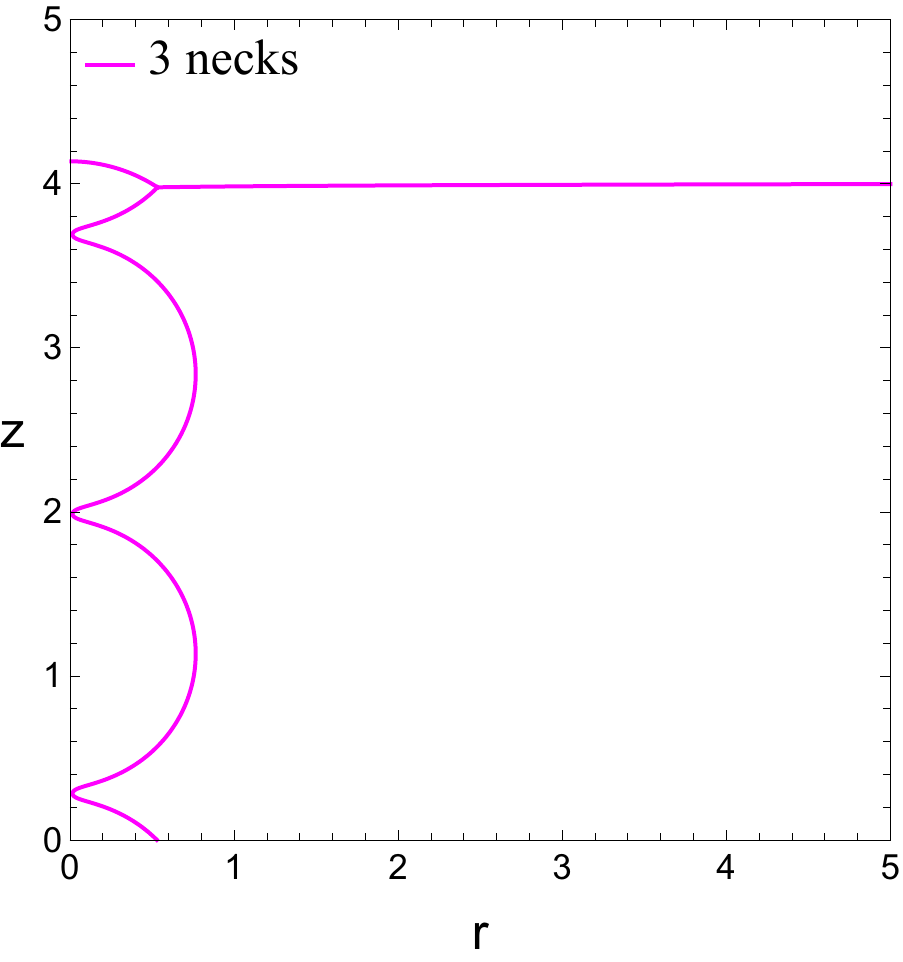}}
\caption{ Effects of different volumes, $V_B$, on the solutions for $\eta=0.8$, $\zeta=0.2$ and $\theta=45^{\circ}$. (a) $h_B=0.5082$, (b) $h_B=1.0119$, (c) $h_B=2.9798$, (d) $h_B=4.0054$.}
\label{fig:shapes}
\end{figure}

In Fig.~\ref{fig:shapes}, we show the shape of the bridge equilibrium solution for four different values of $h_B$. For a relatively small $h_B$, such as $h_B=0.5082$ in Fig.~\ref{fig:shapes}(a),  we obtain a solution with a smooth curve~$2$, connecting the solid bottom with the triple point. Instead, for larger values of $h_B$, this curve is no longer smooth, but develops one or more \emph{necks} (Fig.~\ref{fig:shapes}(b)-(d)). More specifically, the $h_B$--intervals for which each type of solution is found are:
\begin{eqnarray}
\textit{no neck}: \quad 0.078 &\leq h_B &\leq 2.838,\nonumber\\
1 \,\,\textit{neck}\,:  \quad 0.656 &\leq h_B &\leq 2.449,\nonumber \\
2 \,\textit{necks}:  \quad 2.449 &\leq h_B &\leq 3.510,\\
3 \,\textit{necks}:  \quad 3.510 &\leq h_B &\leq 4.589.\nonumber
\end{eqnarray}
The minimum value of $h_B=0.078$ corresponds to the case where $V_B$ is so small that it leads to $h_w=0$. 

\begin{figure}[ht]
\subfigure [] {\includegraphics[width=0.45\linewidth]{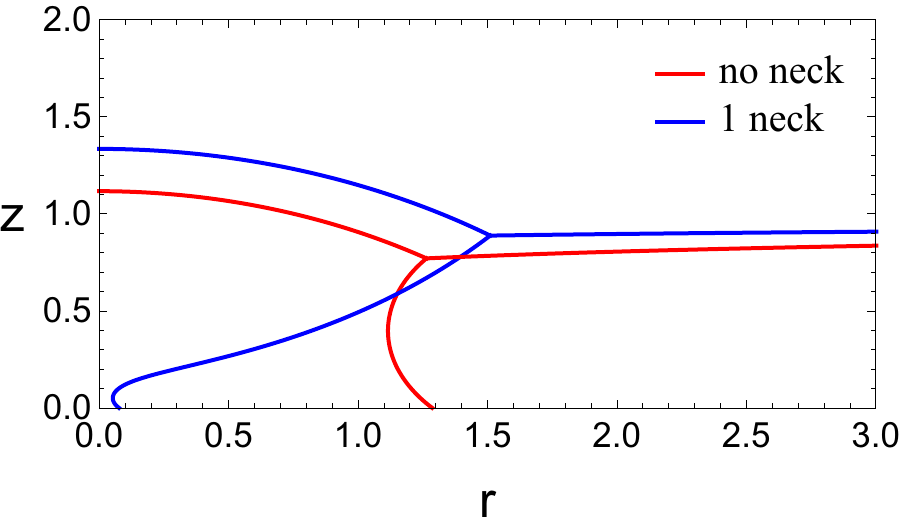}}
\subfigure [] {\includegraphics[width=0.45\linewidth]{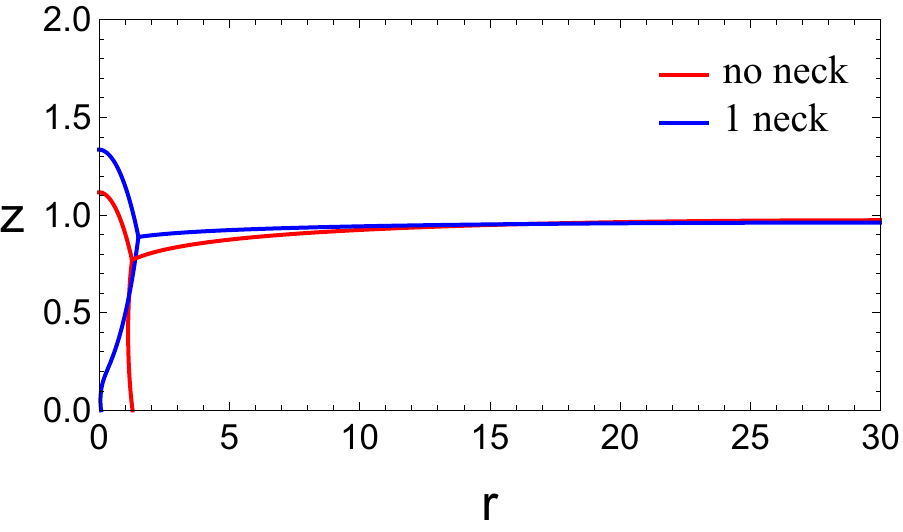}}
\caption{ (a) Two different equilibrium solutions obtained for the same set of physical parameters, ${\cal F}=(\eta,\zeta,\theta, h_B)=(0.8,0.2,45^{\circ},0.9526)$. The initial sets of values for the numerical procedure are obtained with $\phi_0=3 \times 10^{-4}$ for no--neck solution  and with $\phi_0=0.9$ for $1$--neck solution. (b) Same solutions for the whole range of $r$, up to $r_w=30$.}
\label{fig:perfilesS1S2}
\end{figure}

Interestingly, there are two $h_B$--intervals where two types of solutions coexist: i) $(0.656,2.449)$ for no--neck and $1$--neck solutions, and ii) $(2.449,2.838)$ for no--neck and $2$--necks solutions. We present in Fig.~\ref{fig:perfilesS1S2}(a) a typical result with two different types of solutions obtained for $h_B=0.95$. The two solutions were obtained using the same procedure described in Sec.~\ref{sec:NumProc}, but with different initial values of $\phi$. In the case of \emph{no--neck} solution, we used $\phi_0=3 \times 10^{-4}$, while for the \emph{$1$--neck} solution we took $\phi_0=0.9$. Note that the former corresponds to \emph{approximately} a drop plus a tiny lens, while the latter to a lens plus a tiny drop. It can be seen in Fig.~\ref{fig:perfilesS1S2}(a) that, for the solution with $1$--neck (blue line), the position of the contact line at the substrate, $r_s$, is much smaller and very close to the symmetry axis in comparison with the no--neck solution (red line). Instead, $r_s$ is far from the axis and close to the radius of the triple point, $r_\ast$. Note that both solutions are practically coincident for very large $r$, as seen in Fig.~\ref{fig:perfilesS1S2}(b).

The profiles with one or more necks (Fig.~\ref{fig:shapes}(b)-(d)) arise the possibility that these  static solutions might not be stable, but they could break up at their necks and lead to the formaction of separated volumes of liquid $A$. Therefore, we will focus on the study of this possibilty by considering both the geometric and energetic feasability of this situation.

\begin{figure}[htb]
\includegraphics[width=0.6\linewidth]{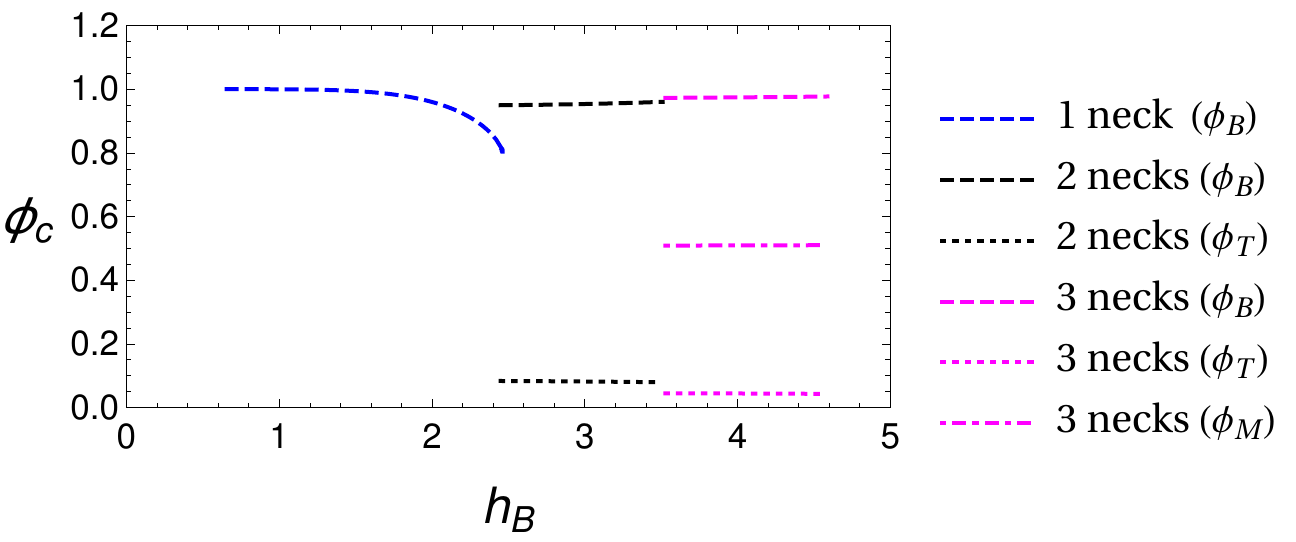}
\caption{Volume fraction of liquid $A$, $\phi_c$, for the upper part of the bridge as function of $h_B$.}
\label{fig:phic_hB}
\end{figure}

If the breakup occurs  \emph{only in one single neck of the solution}, two separate portions of liquid $A$ will be formed, namely, an upper portion of the bridge with volume fraction $\phi_c=V_{upper}/V_A$, and a lower one with a volume fraction $1-\phi_c=V_{lower}/V_A$. Here, the possible values of $\phi_c$ are  determined by the necks positions of a given solution. Since we will consider here no more than $3$--necks solutions, we denote by $\phi_B$ the value of $\phi_c$ for the \emph{bottom} neck (or a single neck when there is only one), by $\phi_T$ that of the \emph{top} neck and by $\phi_M$ that of the \emph{middle} neck. These values depend on $h_B$ and are shown in Fig.~\ref{fig:phic_hB} for each type of solution. Note that $\phi_B$ is very close to unity in the whole range of $h_B$, thus indicating that the breakup at the \emph{bottom} neck will lead to the formation of a large lens and a small drop on the substrate. Instead, $\phi_T$ results very small for both $2$ and $3$--necks solutions, so that a small lens and a large drop will form if the breakup occurs at the \emph{top} neck. Finally, if breakup occurs at the middle neck (which is only possible for three necks), the sizes of both the lens and the drop will be very similar, since $\phi_M \approx 0.5$ within the corresponding $h_B$--interval.

\begin{figure}[ht]
\subfigure[]{\includegraphics[width=0.49\linewidth]{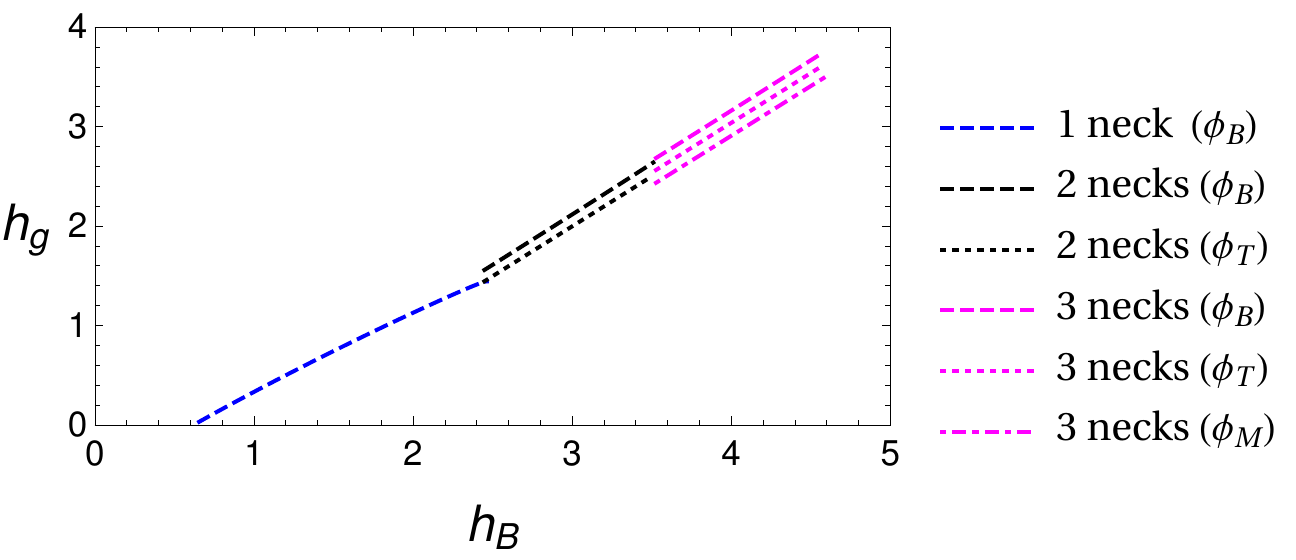}}
\subfigure[]{\includegraphics[width=0.49\linewidth]{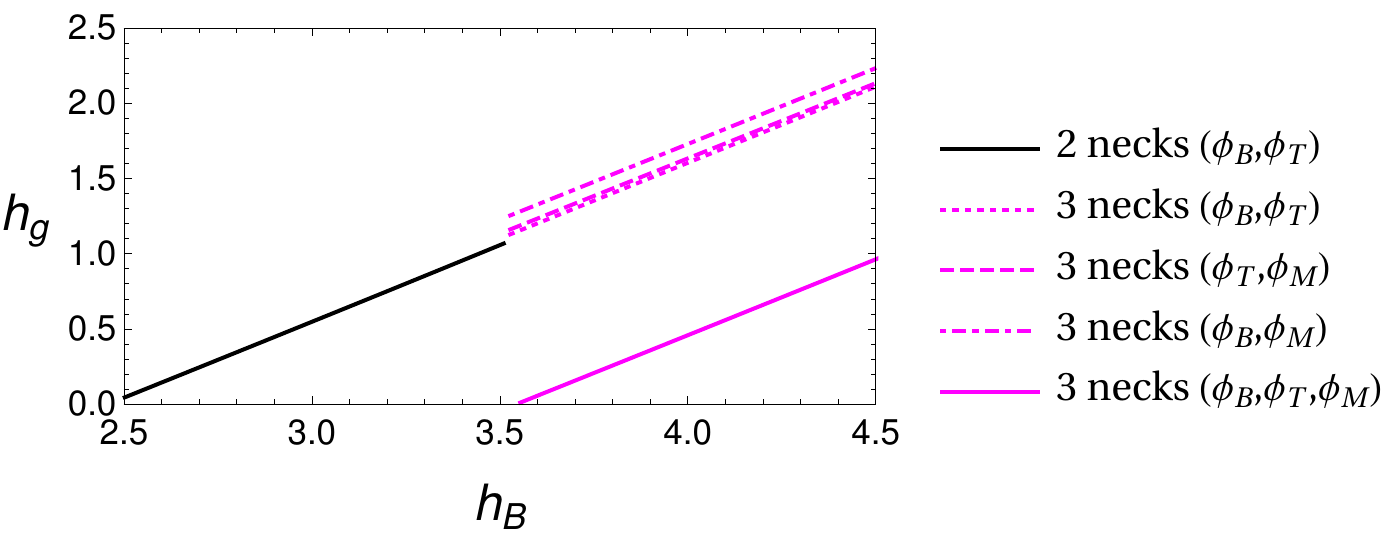}}
\caption{Spatial gap between the bottom of the floating lens and the top of the sessile drop, $h_g$, as a function of $h_B$: (a) For a single breakup of the solution (no matter its number of necks). (b) For more than two simultaneous breakups in two or more necks solutions.}
\label{fig:hg_hB}
\end{figure}

Note that the separated system \emph{lens plus drop} can be formed only if there is a gap of fluid $B$ that precludes any overlaping between both structures when placed on the same axis. It is possible to obtain the separation distance between the bottom of the lens and the top of the drop, $h_g$, from the resulting values of $\phi_c$ for given $h_B$. In Fig.~\ref{fig:hg_hB}(a), we show $h_g$ for the $1$, $2$ and $3$--necks solutions by considering the possible values of $\phi_c$ in each case, namely, $\phi_B$, $\phi_T$ and $\phi_M$ (see upper lines). 

When there is more than one neck, two or three breakups might occur \emph{simultaneously}. In these cases, the separated system is formed by \emph{drop plus lens plus a sphere} for a two breakups (for $2$ and $3$--necks solutions), or by \emph{drop plus lens plus two spheres} for a three breakups (for three and more necks solutions). Thus, the geometrical conditions for the existence of these other configurations is that the sums of the gaps between the separated volumes be positive in each case. This is shown in Fig.~\ref{fig:hg_hB}(b) by the solid lines when all the necks of the solution break up simultaneous and by the dashed lines for simultaneous breakups in pairs for the three neck soluton.  

Therefore, we conclude that the above mentioned breakups are geometrically feasible, since we always have $h_g>0$ for the considered $h_B$--interval, no matter the number of necks of the solution neither their combination of breakups. Note that the beginning of each $h_B$--interval coincides with the condition $h_g=0$ when the number of simultaneous breakups is equal to that of the necks of the solution. In the following, we will explore the solutions and their possible breakups from an energetic point of view.

In order to determine the final state reached by a real system, we calculate the surface energy of each equilibrium solution as well as that of the possible separated systems (drop plus lens plus eventual spherical drops). Thus, we expect that the system will evolve to the lower energy state if the geometrical and topological constraints allow it.

Since only surface energies are taken into account in this problem, their calculation requires the knowledge of the corresponding interfacial tension of each interface, $\sigma_i$, and their area, $S_i$ ($i=1,2,3$). However, the surface energy of the substrate in contact with liquids $A$ and $B$, ${\cal E}_s$, is also part of the total energy, ${\cal E}$, and must be taken into account in the calculation for a given configuration. Thus, we have
\begin{equation}
{\cal E}= {\cal E}_1 + {\cal E}_2 + {\cal E}_3 + {\cal E}_s,
\end{equation}
where ${\cal E}_s$ can be written as
\begin{equation}
{\cal E}_s =  \pi \left[ \sigma_{sA} \, r_s^2 + \sigma_{sB} \, \left( r_w^2 - r_s^2 \right) \right] {\cal R}_0^2.
\end{equation}
Here, $\sigma_{sA}$ and $\sigma_{sB}$ are the surface tensions between the substrate and liquids $A$ and $B$, respectively, and are usually unknown quantities. Therefore, we use the Young's equation, which determines the contact angle $\theta$ as 
\begin{equation}
\cos \theta = \frac{\sigma_{sB}-\sigma_{sA}}{\sigma_2},
\end{equation}
to eliminate $\sigma_{sA}$ from ${\cal E}_s$. We obtain
\begin{equation}
{\cal E}_s = - \sigma_2 \cos \theta \,\pi r_s^2 {\cal R}_0^2 +{\cal E}_s^0,
\end{equation}
where ${\cal E}_s^0=\sigma_{sB} \,\pi r_w^2 {\cal R}_0^2$ is the constant surface energy of the interface substrate--liquid $B$ when the liquid $A$ is absent. Note that ${\cal E}_s^0$ is an irrelevant energy since $r_w$ is arbirarily large and, consequently, the same observation can be made on ${\cal E}_3$. Therefore, we will consider only the difference ${\cal E}_3-{\cal E}_3^0$, where ${\cal E}_3^0$ is the corresponding value of ${\cal E}_3$ for a completely flat surface, i.e., 
\begin{equation}
{\cal E}_{3}^0 = \sigma_{3} \, \pi r_{w}^2 {\cal R}_0^2= \sigma_{ref} \left( 1+\eta-2 \zeta \right)\, \pi r_{w}^2 {\cal R}_0^2.
\label{eq:energyS3}
\end{equation}
In summary, the relevant system energy is given by
\begin{equation}
{\cal E}^\prime = {\cal E} - {\cal E}_s^0 - {\cal E}_3^0,
\end{equation}
and it is independent of $r_w$. Note that ${\cal E}_s^0 + {\cal E}_3^0$ is the system energy without the liquid volume $V_A$.

In order to define a dimensionless energy of order of unity, we consider the energy of a single spherical drop of volume $V_A$ resting on the substrate and completely covered by liquid $B$, i.e.
\begin{equation}
{\cal E}_d=\pi r_{d,0}^2 \sigma_2 {\cal R}_0^2 \left[ \sec^2 \frac{\theta}{2}-\cos \theta \right] + {\cal E}_s^0 + {\cal E}_3^0,
\end{equation}
where $r_{d,0}$ is given by Eq.~(\ref{eq:rd}) for $\phi=0$. Therefore, we take the reference energy 
\begin{equation}
{\cal E}_{ref}^\prime = {\cal E}_d - {\cal E}_s^0 - {\cal E}_3^0,
\end{equation}
so that we define the \emph{dimensionless} total energy as 
\begin{equation}
E=\frac{{\cal E}^\prime}{{\cal E}_{ref}^\prime}.
\label{eq:energyScale}
\end{equation}

\begin{figure}[htb]
\subfigure{\includegraphics[width=0.9\linewidth]{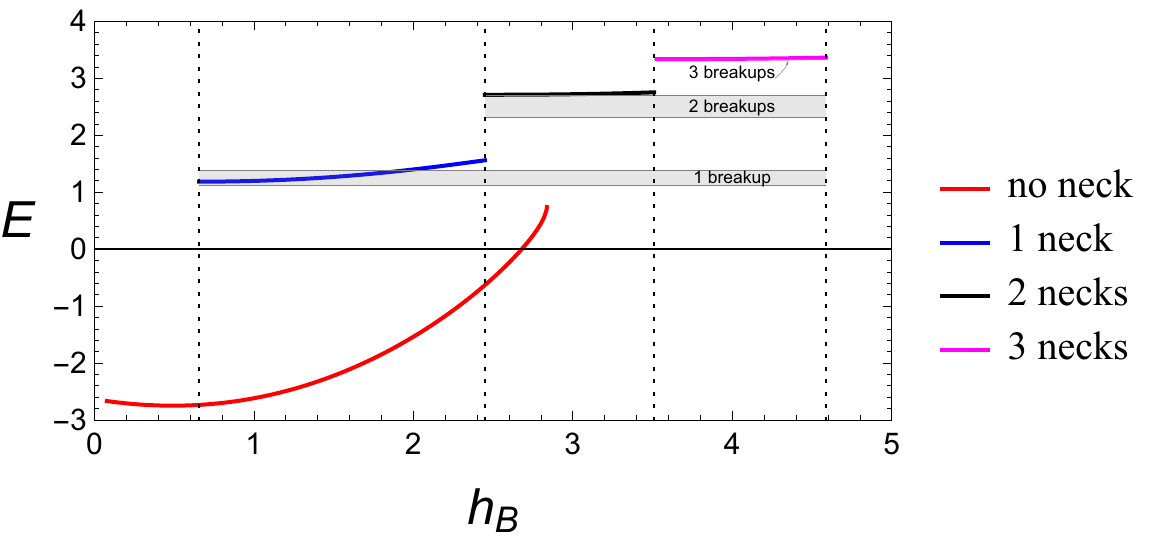}}
\caption{Total energy, $E$, of each type of solution presented in Fig.~\ref{fig:shapes} as a function of the equivalent thickness of liquid $B$, $h_B$. The vertical dotted lines divide the regions of existence for each type of solution. Note that the \emph{smooth bridge} solution is the only one that coexist with one and two \emph{necks} solutions.}
\label{fig:EBridhb}
\end{figure}

By using these definitions, we calculate the energy of the four types of solutions shown in Fig.~\ref{fig:shapes} for their corresponding $h_B$ intervals (see Fig.~\ref{fig:EBridhb}). We observe that the \emph{no--neck} solution is the one that has the lowest energy, being negative for $h_B \lesssim 2.68$; this means that its energy can be even smaller than that of the system without liquid $A$. Moreover, the energy is minimum for $h_B \approx 0.49$.

As regards to the solutions with necks, their energies are monotonous increasing functions of $h_B$ within each interval with positive jumps at the endings. The shadowed regions in Fig.~\ref{fig:EBridhb} indicate the energy ranges of the resulting separated systems when there are  one, two or three breakups at the necks. The higher is the number of breakups, the higher are the corresponding energies, so that the single breakup is the most likely to occur. 

\begin{figure}[htb]
\includegraphics[width=0.9\linewidth]{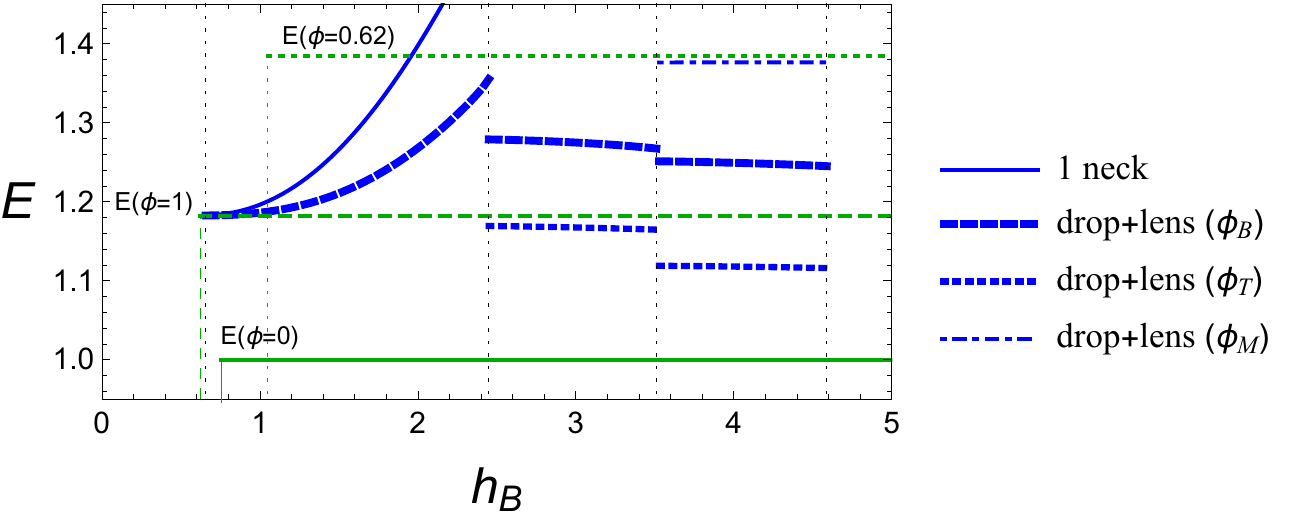}
\caption{Closeup of Fig.~\ref{fig:EBridhb} for $1$ breakup region. The solid blue line stands for one neck solution, also shown in Fig.~\ref{fig:EBridhb}, is plotted for completeness. The dashed lines correspond to the energy, $E$, of the separated system \emph{lens plus drop} when the breakup occurs at the bottom (thick dashed--lnes), at the top (dotted lines) and at the middle (dot--dashed lines).  The horizontal lines correspond to the drop energy without lens, $E(\phi=0)$ (green solid line), the lens energy without drop, $E(\phi=1)$ (green dashed line) and the maximum energy of the \emph{lens plus drop} system, $E(\phi=0.62)$ (green dotted line).}
\label{fig:E_hb}
\end{figure}

A closeup of the region for $1$ breakup is shown in Fig.~\ref{fig:E_hb}, where the possibilities for all three solutions can be analyzed in more detail. The solid blue line corresponds to the $1$--neck solution, also shown in Fig.~\ref{fig:EBridhb}. In principle, each type of solution can have as many breakups as its number of necks, so that the resulting separated systems consist of a \emph{drop plus lens}. If the breakup occurs at the bottom neck, thus determining the volume fraction $\phi_B$, the corresponding energies are represented by the thick dashed lines for $1$, $2$ and $3$--necks solutions. Analogous description applies to the breakups at top ($\phi_T$, for two and three necks solutions) and middle ($\phi_M$, for the three necks solution). 

\begin{figure}[htb]
\includegraphics[width=0.5\linewidth]{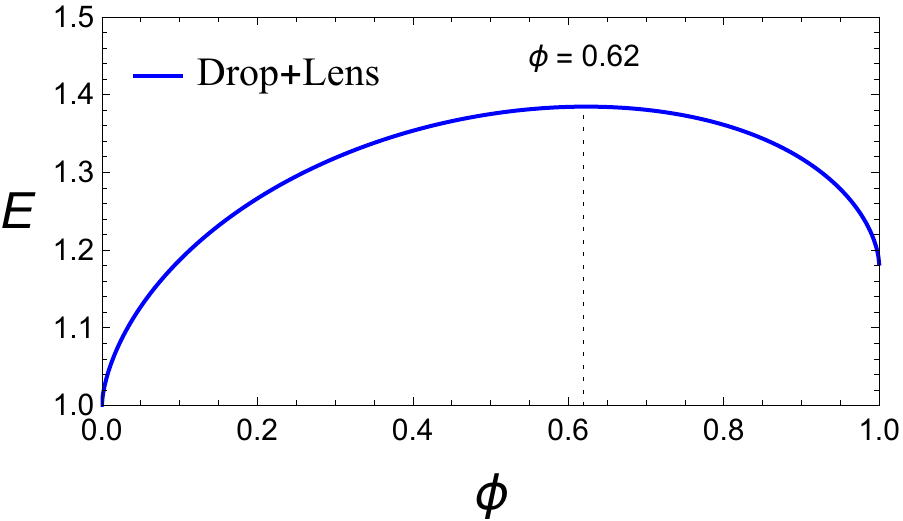}
\caption{Energy of the \emph{lens + drop} configuration as function of the volume fraction $\phi$.}
\label{fig:fragEn}
\end{figure}

In order to visualize the relative sizes of the resulting \emph{drop plus lens} system, we relate these energies with the values of the volume fractions. To do so, in Fig.~\ref{fig:fragEn} we show the energy of the \emph{lens plus drop} system as a function of the volume fraction, $\phi$, for $\theta=45^\circ$. Interestingly, $E(\phi)$ is not a monotonous function, but it presents a maximum at $\phi=0.62$ where $E=1.384$. The minimum value is reached at $\phi=0$ where $E=1$, and the system is formed only by the drop. At $\phi=1$ (only lens), the energy adopts an intermediate value, $E=1.182$. 
Consequently, when the breakup occurs at the bottom, the value of $\phi_B$ is relatively large ($>0.62$), so that the volume lens is much greater than that of the drop. Instead, when the breakup occurs at the top, $\phi_T$ is relatively small ($<0.62$), and the separated system consists of a small lens and a large drop. Finally, when the breakup occurs at the middle (in the three necks solution), $\phi_M$ is close to $0.6$, so that both the drop and lens have approximately similar volumes.
 
\section{Effects of the contact angle, $\theta$}
\label{sec:ContactAngle}

\begin{figure}[ht]
\subfigure [$\,\theta=30^\circ$] {\includegraphics[width=0.45\linewidth]{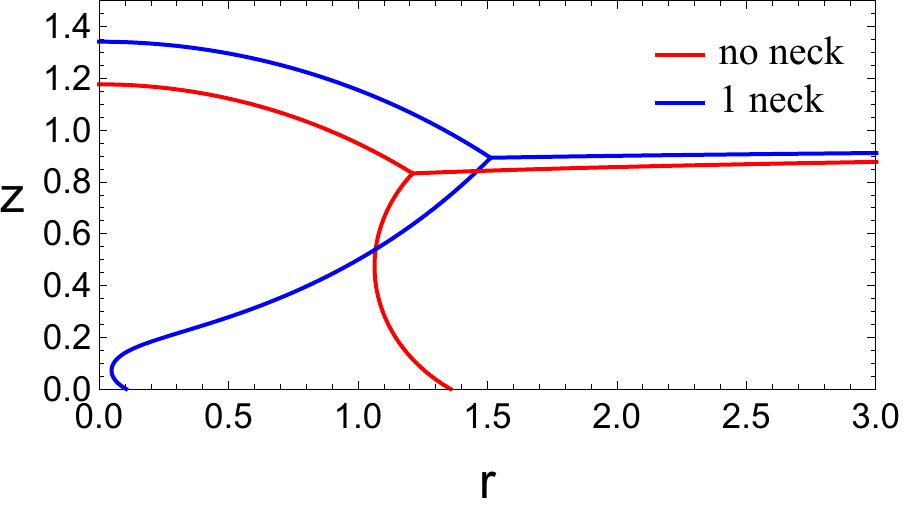}}
\subfigure [$\,\theta=60^\circ$] {\includegraphics[width=0.45\linewidth]{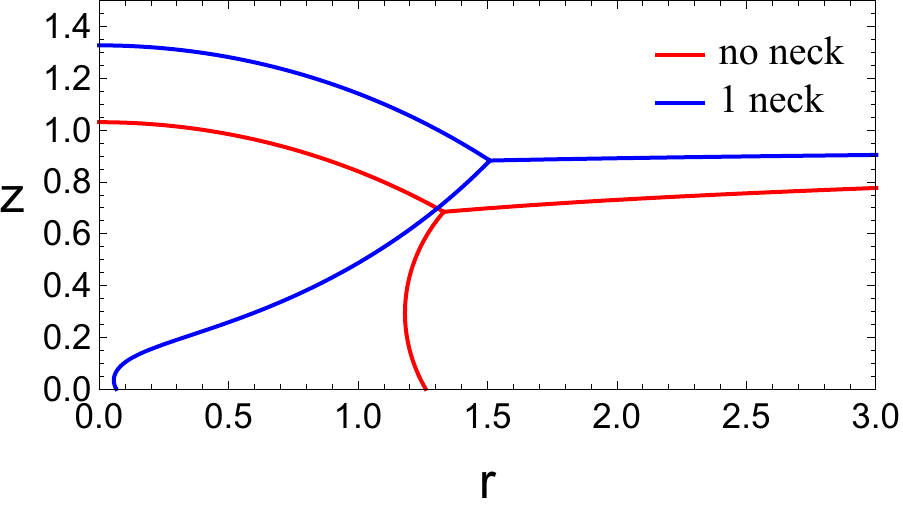}}
\subfigure [$\,\theta=90^\circ$] {\includegraphics[width=0.45\linewidth]{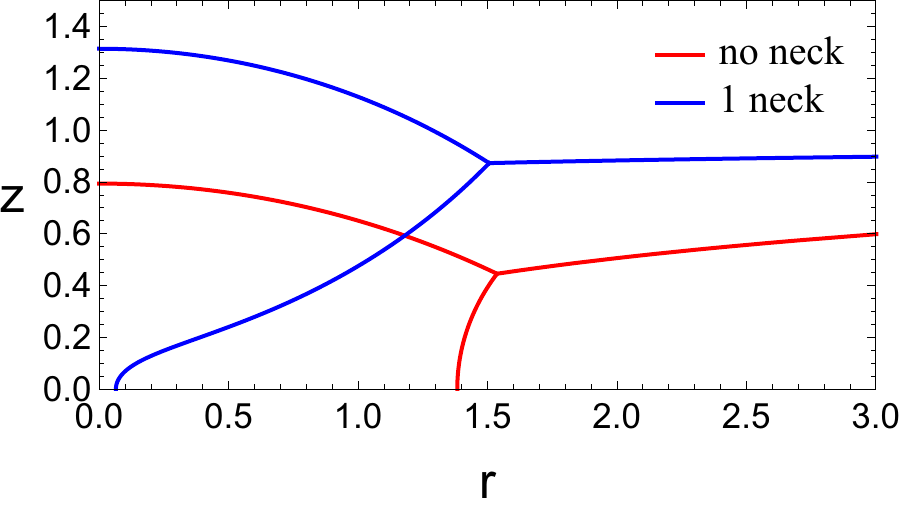}}
\subfigure [$\,\theta=120^\circ$] {\includegraphics[width=0.45\linewidth]{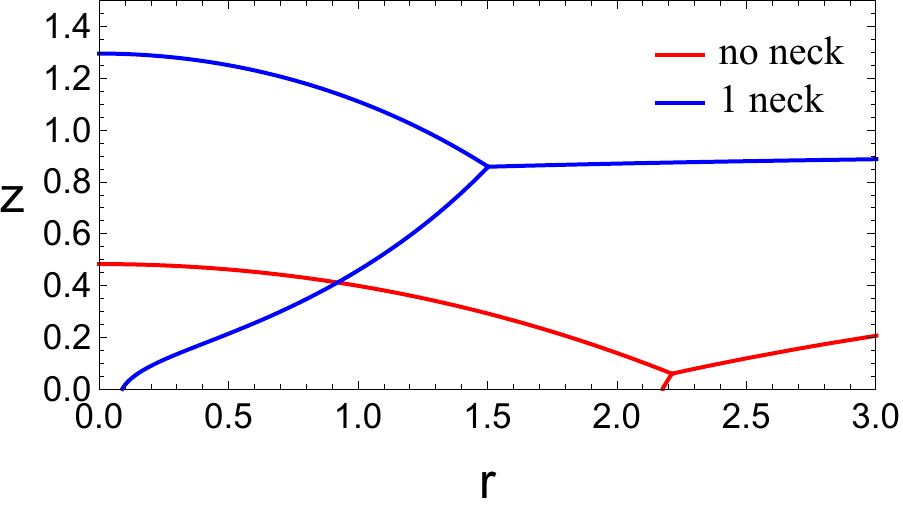}}
\caption{ Effects of different contact angles on the solutions for $\eta=0.8$, $\zeta=0.2$ and $h_B=0.9526$. (a) $\theta=30^\circ$, (b) $\theta=60^\circ$, (c) $\theta=90^\circ$, (d) $\theta=120^\circ$.}
\label{fig:thetas}
\end{figure}

The value of the contact angle, $\theta$, can be modified by changing the type of substrate at the bottom of the vessel. Here, we keep $(\eta,\zeta,h_B)=(0.8,0.2,0.9526)$ and we vary $\theta$. In Fig.~\ref{fig:thetas} we show the solutions for $\theta=30^\circ, 60^\circ, 90^\circ, 120^\circ$ (see Fig.~\ref{fig:perfilesS1S2}(a) for $\theta = 45^\circ$). We observe that, for all these five values of $\theta$, we obtain only no--neck and $1$--neck solutions. As $\theta$ increases, the no~neck solutions (red lines) become for more flatten (both $h_1$ and $z_\ast$ diminish) and more extended ($r_s$ increases). Conversely, the $1$--neck solutions (blue lines) do not have significant changes for different $\theta$'s, except at the region in contact with the substrate. 

\begin{figure}[ht]
\subfigure{\includegraphics[width=0.49\linewidth]{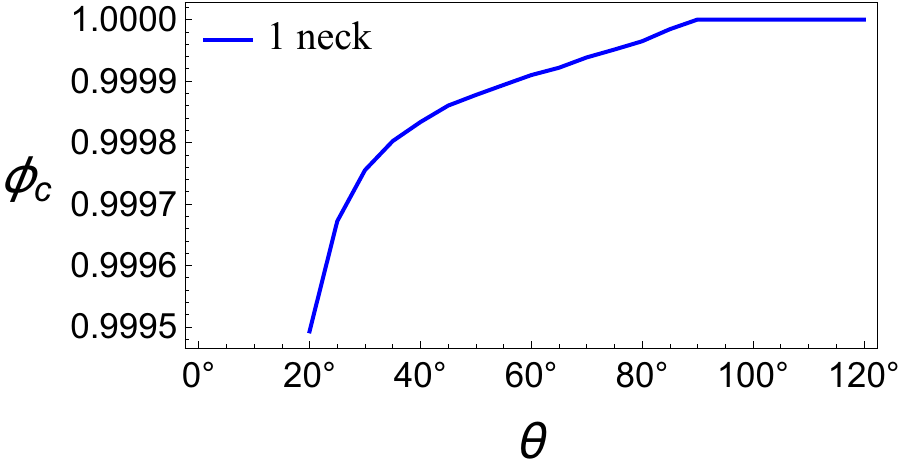}}
\subfigure{\includegraphics[width=0.49\linewidth]{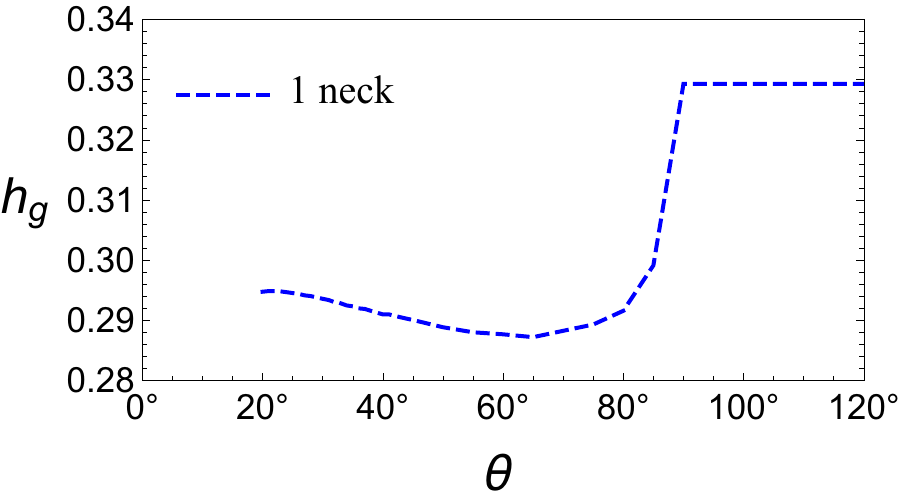}}
\caption{(a) Volume fraction of liquid $A$, $\phi_c$, for the upper part of the bridge as function of $\theta$. (b) Spatial gap between the bottom of the floating lens and the top of the sessile drop, $h_g$, as a function of $\theta$.}
\label{fig:phic_hg_theta}
\end{figure}

As a consequence, the corresponding volume fraction, $\phi_c$, remains practically constant and very close to unity (see Fig.~\ref{fig:phic_hg_theta}(a)). In Fig.~\ref{fig:phic_hg_theta}(b), we show that the gap distance of the separated solution is positive and around $0.3$ for the whole $\theta$--range. Therefore, the breakup of the neck in the $1$--neck solution will lead to a large lens plus a tiny droplet on the substrate.

\begin{figure}[ht]
\subfigure [$\,\theta=90^\circ$] {\includegraphics[width=0.45\linewidth]{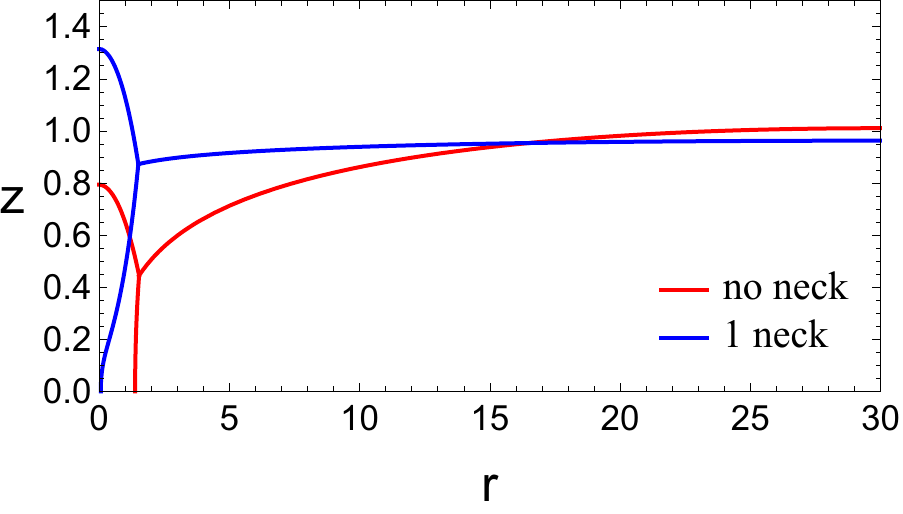}}
\subfigure [$\,\theta=120^\circ$] {\includegraphics[width=0.45\linewidth]{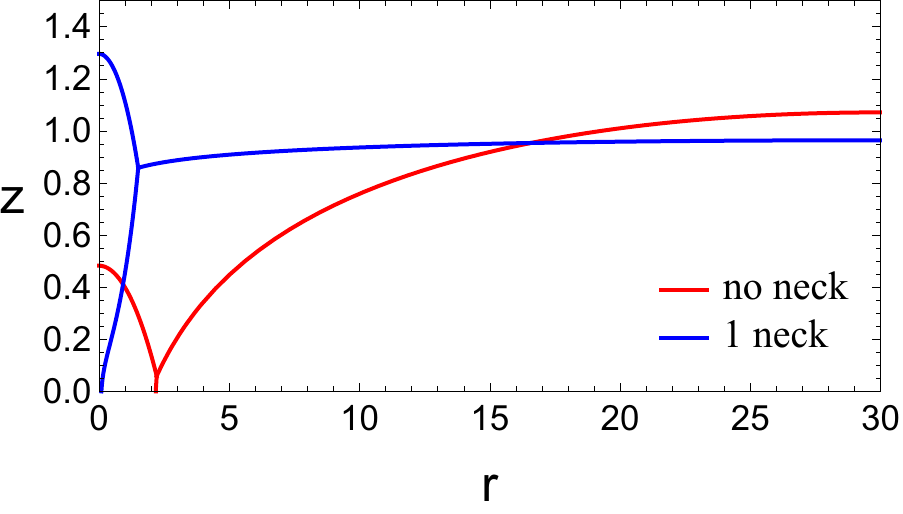}}
\caption{Solution for $\theta >90$ and large $r$ for $\eta=0.8$, $\zeta=0.2$ and $h_B=0.9526$. (a) $\theta=90^\circ$, (b) $\theta=120^\circ$.}
\label{fig:large_theta}
\end{figure}

Note that for large $r$ both solutions are almost coincident for $\theta <90^\circ$ (see e.g.,  Fig.~\ref{fig:perfilesS1S2}(b) for $\theta=45^\circ$). However, the curves~$3$ of both solutions are very different for larger values of $\theta$ (see Fig.~\ref{fig:large_theta}). Both the increase of $r_s$ and the decrease of $h_1$, also represented by the displacement of the triple contact point ($r_\ast$, $z_\ast$) to larger $r$ and lower $z$, imply a significant change of curve~$3$ with the consequent variation of $h_w$ (due to the constancy of $h_B$).

\begin{figure}[htb]
\subfigure[]{\includegraphics[width=0.45\linewidth]{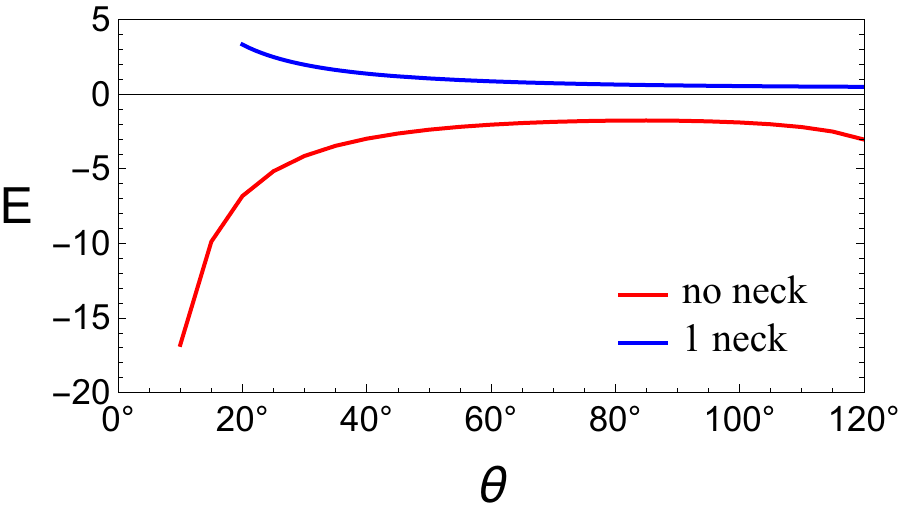}}
\subfigure[]{\includegraphics[width=0.45\linewidth]{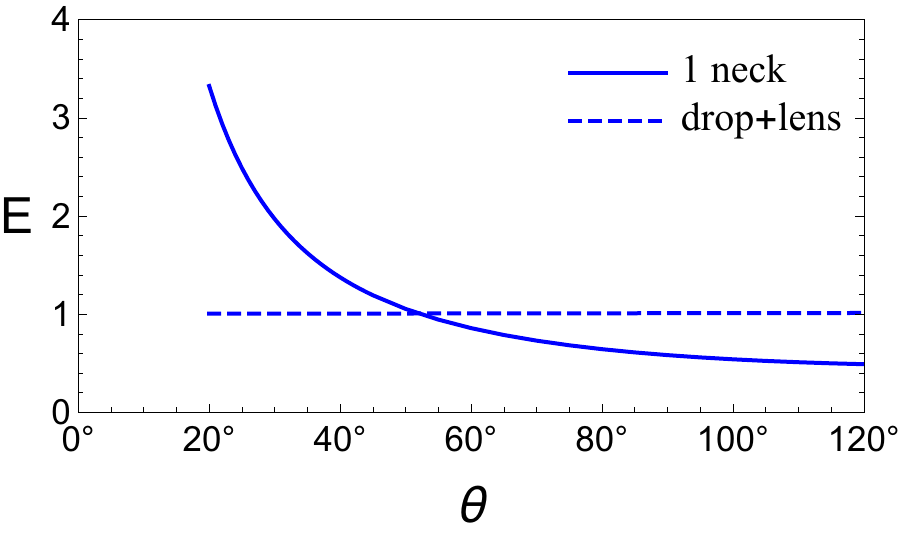}}
\caption{(a) Total energy, $E$, of the no--neck and $1$--neck solutions for $h_B=0.9526$ as a function of the contact angle, $\theta$. (b) Total energy, $E$, of the $1$--neck solution in (a) (solid line) and the corresponding energy of the separated system \emph{lens plus drop} (dashed lines).}
\label{fig:E_theta}
\end{figure}

As regards to the energetic analysis for variable $\theta$ (see Fig.~\ref{fig:E_theta}(a)), we observe that the energy of the no--neck solution remains negative for all $\theta$--range, while that of the $1$--neck solution is positive and monotonically decreases with $\theta$. Interestingly, the corresponding separated system \emph{lens plus drop} for the breakup of the $1$--neck solution has higher energy than the continuous solution for $\theta >52^\circ$ (see Fig.~\ref{fig:E_theta}(b)). Therefore, one expects the breakup of the solution only for smaller $\theta$'s. 

\section{Effects of the surface tensions ratii}
\label{sec:eta}
In this section, we analyze the effects related to the variation of $\eta$, while keeping fixed the other parameters, namely, $(\zeta,\theta,h_B)=(0.2,45^\circ,0.9526)$. As mentioned above, a similar variation of $\zeta$ is not signficative, since both $\eta$ and $\zeta$ only modify the ratii of surface tensions (see Eq.~(\ref{eq:eta_zeta})). Here, we consider $0.5 \leq \eta=\sigma_2/\sigma_1 \le1 1.8$, so that $1.1 \leq \sigma_3/\sigma_1 \leq 2.4$.

\begin{figure}[htb]
\subfigure[]{\includegraphics[width=0.45\linewidth]{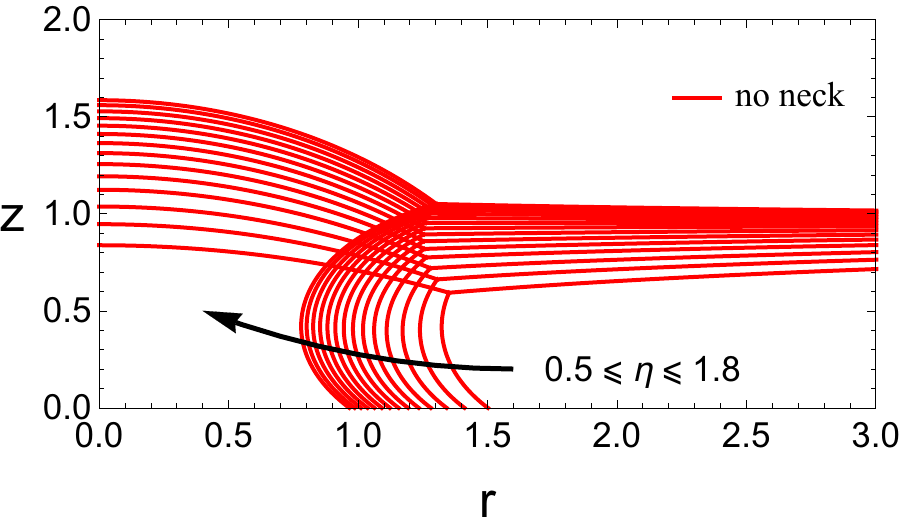}}
\subfigure[]{\includegraphics[width=0.45\linewidth]{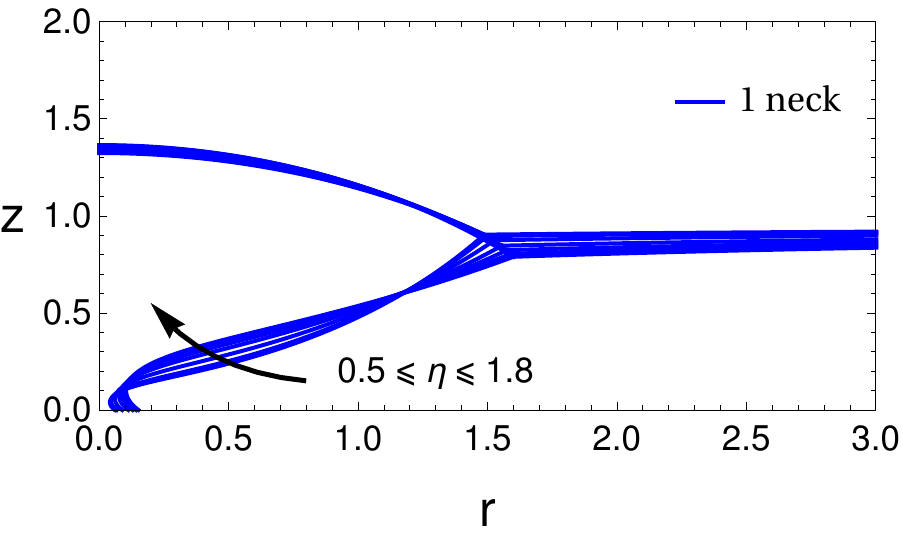}}
\caption{Effects of varying $\eta$ for: (a) the no--neck solutions, (b) the $1$--neck solutions. Here, we used $(\zeta,\theta,h_B)=(0.2,45^\circ,0.9526)$.}
\label{fig:perf_eta}
\end{figure}

The solutions obtained in this $\eta$--range are shown in Fig.~\ref{fig:perf_eta} for both no--neck and $1$--neck cases. As regards to the $1$--neck solutions, their volume fractions at the neck are $\phi_c \approx 0.99$, while those of the gap distances for the separated systems \emph{drop plus lens} are $h_g \approx0.45$. Therefore, different ratii of surface tensions have no significative effects on the main properties of the solutions. 

\section{Summary and Conclusions}
\label{sec:conclu}

In this work, we obtain families of solutions for the problem of a fluid volume (${\cal V}_A$) in contact with a solid substrate and partially covered by another inmiscible fluid (${\cal V}_B$), while a third one (${\cal V}_C$) completely covers both of them. These solutions are obtained by means of an iterative numerical scheme that solves all six coupled non--linear second order ODE's with the corresponding twelve boundary conditions. Additional constraints come from the Neumann's equilibrium equations, the matching of all three curves at the triple point and the volumes conservations. Consequently, there are nine unknown constants to be consistenly determined. The guess values of these nine parameters, needed for the iterative method to find the solution curves, are obtained from the separated system \emph{drop plus lens}, whose solution is known analytically. Finally, the solution is defined by four given values of the parameters that determine: i) the ratii of the surface tensions ($\eta$ and $\zeta$), ii) the contact angle at the solid substrate ($\theta$), and iii) the volume ratio ${\cal V}_B/{\cal V}_A$ (or the effective thickness $h_B$). A parametric study of the solution is presented by varying these system properties.

There are two main results of this parametric study: i) there is region of the parameter $h_B$, which stands for the amount of fluid $B$, where we find two qualitatively different equilibrium solutions, namely, one with no--neck and another with $1$--neck. We also find a short $h_B$--interval where both  no--neck and $2$--necks solutions coexist. ii) For $h_B$ beyond these intervals, we find consecutive $h_B$--regions with $2$, $3$ and more number of necks.

Since the width of the necks is very narrow, the question of their eventual breakup naturally arises, i.e., the stability of the equilibrium solutions with necks can be a relevant issue to be taken into account. By means of an energy analysis of the solutions, we study the possibility of the breakups and the eventual formation of a separated system \emph{drop plus lens}, as well as the formation of additional spherical droplets when the breakups occur at more than one neck (for the corresponding solutions). This singular property of different number necks depends only on the value of $h_B$, i.e., the effective height of the external fluid $B$. Interestingly, the variation of other parameters, such as the contact angle, $\theta$, or the ratios of surface tensions ($\eta$ or $\zeta$) do not affect the number of necks, which is only determined by the chosen value of $h_B$. 

We have proved that several equilibrium solutions are possible and, as a preliminary study, we have considered the energy differences among them. Although this could be used as rule of thumb to indicate the most probable transitions, further dynamical studies are required to ascertain the actual unstable properties of each equilibrium. 

\section*{ACKNOWLEDGMENTS}
The authors acknowledge support from Consejo Nacional de Investigaciones Científicas y Técnicas
(CONICET, Argentina) with Grant PIP 02114-CO/2021 and Agencia Nacional de Promoción Científica y Tecnológica (ANPCyT,Argentina) with Grant PICT 02119/2020. The authors gratefully acknowledge the initial suggestions made by Prof. Howard Stone.

\bibliographystyle{unsrt}
\bibliography{liquidliquid.bib}

\end{document}